\documentclass[letterpaper,twocolumn,10pt]{article}
\usepackage{usenix2019_v3}

\usepackage{tikz}
\usepackage{amsmath}

\usepackage{cite}
\usepackage{multirow}
\usepackage{graphicx}
\usepackage{latexsym}
\usepackage{booktabs}
\usepackage{amssymb}
\usepackage{xcolor}
\usepackage{footnote}
\usepackage{xurl}
\usepackage{arydshln}
\usepackage{longtable}

\usepackage[final]{changes}

\begin{document}

\date{}

\title{\Large \bf How WEIRD is Usable Privacy and Security Research?\\(Extended Version)}

\author{
{\rm Ayako A. Hasegawa}\\
NICT
\and
{\rm Daisuke Inoue}\\
NICT
\and
{\rm Mitsuaki Akiyama}\\
NTT
} 

\maketitle

\begin{abstract}
In human factor fields such as human-computer interaction (HCI) and psychology, researchers have been concerned that participants mostly come from WEIRD (Western, Educated, Industrialized, Rich, and Democratic) countries.
This WEIRD skew may hinder understanding of diverse populations and their cultural differences. 
The usable privacy and security (UPS) field has inherited many research methodologies from research on human factor fields.
We conducted a literature review to understand the extent to which participant samples in UPS papers were from WEIRD countries and the characteristics of the methodologies and research topics in each user study recruiting Western or non-Western participants. 
We found that the skew toward WEIRD countries in UPS is greater than that in HCI.
Geographic and linguistic barriers in the study methods and recruitment methods may cause researchers to conduct user studies locally.
In addition, many papers did not report participant demographics, which could hinder the replication of the reported studies, leading to low reproducibility.
To improve geographic diversity, we provide the suggestions including facilitate replication studies, address geographic and linguistic issues of study/recruitment methods, and facilitate research on the topics for non-WEIRD populations.
\end{abstract}

\section{Introduction}

\renewcommand{\thefootnote}{\fnsymbol{footnote}}
\footnote[0]{This paper is the extended version of the paper presented at
USENIX SECURITY 2024.}
\renewcommand{\thefootnote}{\arabic{footnote}}

Usable privacy and security (UPS) research focuses on the perspective of those who use technology (i.e., humans), which has not been addressed in conventional security research.
It contributes to solving security and privacy issues that arise in the gap between social requirements and technical feasibility, i.e., \textit{social-technical gap}~\cite{10.1207/S15327051HCI1523_5}.
The UPS field has accumulated approximately a quarter of a century of research history. 
UPS papers have been published in various conferences and journals both inside and outside the field of security and privacy, and the number of published UPS papers is increasing every year.
In 1996, Zurko and Simon introduced the concept of UPS, which was called \textit{user-centered security} in their paper. 
Two pioneering UPS papers were subsequently published in 1999.
Adams and Sasse advocated that human factors should be considered in the design of security mechanisms~\cite{10.1145/322796.322806}.
Further, Whitten and Tygar pointed out a lack of good user interface design for security and applied several usability testing approaches with signing and encrypting an email message~\cite{271563}.

The research methodologies, i.e., recruiting participants and conducting a user study, for UPS are often interdisciplinary. 
Many of these methodologies are drawn from human-computer interaction (HCI), psychology, and behavioral sciences.
In these research fields, the majority of participants have been reported to be Western, which has caused the issue that many of the findings accumulated over several years may not generalize to other countries and cultures.
Henrich et al. indicated that U.S. participants were common outliers on many psychological measures and this fact made these participants ``WEIRD'' (Western, Educated, Industrialized, Rich, and Democratic)~\cite{BBS2010_Henrich}.
Linxen et al.~\cite{CHI2021_Linxen} published a paper in 2021 entitled ``How WEIRD is CHI?'' in which they conducted a literature review and quantitatively confirmed the WEIRD skew of participant samples at CHI, one of the top conferences in HCI.
Because the residents of WEIRD countries are a minority of the world population, this fact is suggested to significantly impact the generalizability of HCI research.
Although the WEIRD framework, which includes indicators based on national statistics or characteristics, has the limitation of assessing participant sampling (e.g., in the Global South individually wealthy people may be more accessible), it has been successful in providing a broad picture of the geographic diversity of the participants.

In UPS, some studies have also identified critical gaps concerning security and privacy between WEIRD and non-WEIRD societies/populations that stem from environmental aspects (e.g., IT resource usage, documentation, and laws/regulations) and personal aspects (e.g.,  (mis)conception, preference, and susceptibility)~\cite{CSCW2017_Ahmed, JGITM2019_Kshetri, CHI2023_Herbert, PETS2017_Li, HAISA2017_Butavicius, SOUPS2018_Sambasivan, EuroUSEC2020_Qahtani, CSCW2021_Li}. 
We further assume that demographic skew may also influence \textit{attack feasibility}. 
Researchers in UPS sometimes leverage user studies to demonstrate the feasibility of attacks, which is a unique approach to UPS. 
These observations suggest that threat assessment, design implications, interventions, and support technologies proposed based on the findings of conventional WEIRD-skewed studies may not be sufficiently beneficial to or actionable in diverse non-WEIRD societies/populations and may instead hinder facilitating security and privacy in such societies/populations.
Even with the harms of the WEIRD skew to UPS research, the degree of geographic breadth and factors that impede geographic diversity have not been sufficiently investigated in UPS.
Moreover, the methodologies used and topics addressed are still either unrevealed or overlooked from the WEIRD perspective.
In this study, we conduct a systematic literature review in UPS to answer the following research questions (RQs).
\smallskip\\
\textbf{RQ1}: \textit{To what extent are participant samples in UPS papers from WEIRD countries? How has this changed over time?}
\smallskip\\
\textbf{RQ2}: \textit{What are the differences in the characteristics of the methodologies and research topics in each user study recruiting Western or non-Western participants?}
\smallskip

We reviewed 715 UPS papers, which recruited participants, from representative security and HCI conferences over a recent five-year period (2017--2021).
We used the WEIRD framework for assessing the geographic diversity of participant samples on the basis of extracted participant information and analyzed the author affiliations, methodologies, and research topics.
We found that a large majority of participant samples in UPS papers were skewed toward WEIRD countries and particularly involved Western participant samples (78.89\%).
This skew is higher than that in the results of Linxen et al.'s study, which reviewed CHI papers (73.13\%). 
Linxen et al. reported that the percentage of non-Western participant samples in CHI papers largely increased (from 16.31\% to 30.24\%, 1.85x) in the past five years, whereas our results revealed that the percentage in UPS papers decreased (25.00\% to 20.77\%, 0.93x).
Authors tend to recruit convenient participants from the countries of their affiliations, e.g., students of the local university who are easily accessible for university researchers.
Because most authors are affiliated with institutions in Western countries, the participant samples are more likely to be skewed toward Westerners. 
Moreover, we identified that a non-negligible number of papers did not report participant demographics. 
This tendency was more pronounced in non-HCI and non-UPS-focused conferences. 
Such papers with low reproducibility may hinder replication studies.
We confirmed that there is a large Western skew in every research topic. 
In particular, attack-feasibility studies are skewed heavily toward Western and suffered from demographic issues: the feasibility of the attacks can be affected by participant demographics due to relying on targets' English proficiency.

We provide the following recommendations for making UPS research less WEIRD.
First, a replication study should be facilitated in UPS because most UPS papers have examined only WEIRD countries and not considered/discussed whether geographic diversity affects the results. 
Second, geographic issues of study and recruiting methods should be addressed.
Third, researchers should also be diversified (e.g., non-Western researchers) to overcome linguistic barriers and communicate precisely and deeply with and understand non-WEIRD populations. 
Finally, we discuss UPS research topics to be addressed in the future.

\section{Background and Related Work}

\subsection{Original Study: How WEIRD is CHI?}
\label{section:related_original}

Linxen et al. published a paper titled ``How WEIRD is CHI?'' in 2021 (hereafter referred to as the Linxen study)~\cite{CHI2021_Linxen}. 
The Linxen study was motivated by the issues raised in psychology: (1) most findings in psychology were based on U.S. undergraduate students~\cite{AP2008_Arnett} and (2) findings drawn from U.S. participants were frequently outliers when compared with those from other countries~\cite{BBS2010_Henrich}.
Then, Linxen et al. investigated the geographic diversity of participants recruited in 2,768 papers published between 2016--2020 at CHI, which is the most prestigious academic conference in the HCI field.
They showed that the participant samples of the CHI papers were highly skewed toward WEIRD countries, e.g., 73.13\% of the participant samples were from Western countries. 
However, encouragingly, they also showed that the percentage of non-Western participant samples had been increasing in recent years.

\subsection{Harms of WEIRD skew to UPS Research}
\label{subsubsection:Harms of WEIRD skew to UPS Research}

Some UPS studies have focused on non-WEIRD societies/populations and identified critical gaps concerning security and privacy between WEIRD and non-WEIRD societies/populations that stem from environmental and individual aspects. 
These results suggest actual and potential harms to UPS research; design implications, interventions, and support technologies that are proposed on the basis of the findings of conventional WEIRD-skewed studies may not be sufficiently beneficial to or actionable in non-WEIRD societies/populations.

\noindent\textbf{Environmental aspects.} 
Gaps have been reported between WEIRD and non-WEIRD societies/populations regarding environmental aspects, e.g., IT resource usage, documentation, and laws/regulations. 
%
%
The first one is the gap in IT resource usage. 
Whereas people in WEIRD societies usually own devices individually, cultural contexts such as economic constraints, social values, and social power relations drive shared device practices in non-WEIRD societies, which create new challenges to ensuring individual privacy~\cite{CSCW2017_Ahmed}.
In particular, several studies have reported that women who live in the Global South, where gender inequality is greater than in Western countries, are likely to be vulnerable to privacy abuse, harassment, and fraud when they use IT resources~\cite{SOUPS2018_Sambasivan, EuroUSEC2020_Qahtani}.
%
%
The second one is the gap in documentation. 
Security and privacy documents (e.g., advice, guidelines, and training materials) notably influence individuals' security and privacy practices~\cite{SP2004_Yan, SOUPS2020_Reinheimer}. 
The U.S., for instance, has issued numerous critical security and privacy documents, e.g., NIST SP-800 series~\cite{NIST_SP800}, which are referenced by other countries. 
However, these documents are much less widely available in non-English-speaking countries than in English-speaking countries due to lower English proficiency levels and the additional cost of translation~\cite{JGITM2019_Kshetri, SOUPS2021_Hasegawa}.
%
%
The third one is the gap in laws/regulations. 
Data privacy laws and regulations play a vital role in protecting individual privacy rights and promoting responsible data practices among service providers. 
Many studies have investigated whether services comply with laws and regulations, e.g., EU GDPR, and the usability of the law-compliant interfaces for user privacy~\cite{CHI2020_Nouwens, CCS2019_Utz}. 
However, in addition to Europe and the U.S., many countries in regions such as Asia, South America, and Africa have already enacted numerous data privacy laws independently~\cite{law_bycountry1, law_bycountry2}, suggesting the anticipated divergence in the treatment of user privacy in services across different countries. 
Moreover, the countries where data privacy laws have not been established have usually been overlooked by the abovementioned study approaches based on existing privacy laws and regulations.

\noindent\textbf{Individual aspects.} 
Gaps have also been reported regarding individual aspects, e.g., (mis)conceptions, preference, and susceptibility. 
It is important to note that individuals and their surrounding environment are not mutually independent entities, as individuals are influenced by their environment~\cite{JGIM2002_Straub, IJIM2019_Li}.
%
%
The first one is the gap in (mis)conceptions. 
Many studies have identified that misconceptions of security and privacy technologies lead to insecure behavior and increase users' risk of being harmed~\cite{SP2017_Abu, PETS2021_Story}. 
However, Herbert et al. identified that one of the most important factors influencing user misconceptions is the country of residence, with greater differences between Western and non-Western countries~\cite{CHI2023_Herbert}.
This result suggests further investigation of the reasons for these differences and how to debunk misconceptions tailored to each country. 
%
%
The second one is the gap in preferences. 
Many studies have identified users' privacy preferences on willingness to share their data with companies and disclose their data on social media and have proposed techniques to recommend privacy settings on the basis of users' privacy preferences~\cite{SOUPS2013_Leon, JCMC2014_Taddicken, TOCHI2015_Watson}. 
Some studies have reported that such privacy preferences vary among countries~\cite{SEC2021_Cao2021, CSCW2021_Li, PETS2017_Li}. 
For example, Li et al. showed that when users were informed that third-party companies were accountable for data collection, users from collectivistic countries (e.g., China) were less likely to share their data, whereas users from individualistic countries (e.g., U.S) were more likely to share their data~\cite{PETS2017_Li}.
%
%
The third one is the gap in susceptibility. 
Many studies have identified attackers' phishing techniques and users' coping strategies and proposed interventions~\cite{SEC2019_Van, SOUPS2021_Wash, SOUPS2021_Franz}. 
However, Butavicius et al. demonstrated that users from individualistic cultures were significantly less likely to be susceptible to phishing emails~\cite{HAISA2017_Butavicius}, suggesting that users in collectivistic countries may be more exposed to phishing threats. 
Moreover, Hasegawa et al. unveiled that non-native English speakers used different coping strategies for suspicious emails in English from suspicious emails in their native language~\cite{SOUPS2021_Hasegawa}.

\smallskip
Motivated by these observations, we focus on the WEIRD perspective in UPS. 
Ensuring security and privacy is not limited to WEIRD societies alone but rather a shared global challenge. 
To strive to equitably provide security and privacy globally, it is imperative to broaden the focus of UPS to include non-WEIRD societies; specifically, understanding the similarities and differences between WEIRD and non-WEIRD societies/populations is crucial for addressing this challenge.

\subsection{Literature Reviews in UPS Research}
\label{section:related_UPS}

Many researchers have conducted literature reviews in terms of ``research topics'' of UPS papers.
Garfinkel and Lipfordand~\cite{SLISPT2014_Garfinkel} summarized the research topics of the whole UPS field.
Other reviews focused on the research topics of specific subfields of UPS, such as privacy- and security-related decision-making~\cite{CSUR2017_Acquisti}, social cybersecurity~\cite{SP2022_Wu}, phishing interventions~\cite{SOUPS2021_Franz}, at-risk users~\cite{SP2022_Warford}, sexual abuse~\cite{SP2022_Obada}, cross-cultural privacy issues in social media~\cite{WWW2013_Ur}, and developer-centered security~\cite{EuroUSEC2019_Tahaei}.

On the other hand, few researchers have conducted literature reviews in terms of the ``methodology'' of UPS papers.
Distler et al. conducted a literature review to understand which methods researchers use and how researchers represent risks to participants~\cite{TOCHI2021_Distler}.
Kaur et al. reviewed and compared the methods of UPS papers on experts and non-experts (end users)~\cite{arXiv2021_Kaur}.
In addition to revealing current research practices, these authors identified methodological and ethical challenges in UPS papers, such as no definition of ethically-acceptable deceptions, lack of using theories for result interpretation, and lack of participant diversity.

\begin{table*}[t] 
\caption{Comparison of our study with prior review literature on user studies.}
\label{table:literature comparison}
\hbox to\hsize{\hfil
\scriptsize{
\begin{tabular}{l|llllllr}\hline
Paper & Field & Population & Geographic Diversity & Research Topic & Conference & Period & \# Papers* \\
\hline\hline
Linxen et al.~\cite{CHI2021_Linxen} & HCI & Any & Analyzed from WEIRD perspective & Not analyzed & CHI & 2016--2020 & 2,768 \\
\hline
Distler et al.~\cite{TOCHI2021_Distler} & UPS & Any & Not analyzed & Analyzed & S\&P, SEC, CCS, SOUPS, CHI & 2014--2018 & 284\\
\hline
Kaur et al.~\cite{arXiv2021_Kaur} & UPS & Experts & Analyzed from W perspective & Analyzed & Tier-1/2 in security~\cite{CSranking}, SOUPS, CHI & 2008--2018 & 48 (+48)**\\
\hline
\multirow{2}{*}{Our study} & \multirow{2}{*}{UPS} & \multirow{2}{*}{Any} & \multirow{2}{*}{Analyzed from WEIRD perspective}  & \multirow{2}{*}{Analyzed} & S\&P, SEC, CCS, NDSS, PETS, CHI,   & \multirow{2}{*}{2017--2021} & \multirow{2}{*}{715}\\
     &     &     &        &    & CSCW, SOUPS, USEC, EuroUSEC& & \\
\hline
\end{tabular}}\hfil}
\footnotesize{\hspace{1em} *The number of fully-reviewed papers. \hspace{0.2em} **They reviewed 48 papers on experts plus 48 papers on non-experts for comparison.}
\end{table*}

Table~\ref{table:literature comparison} compares our study with the closely related literature reviews.
Our study is a \textit{quasi}-replication of the Linxen study~\cite{CHI2021_Linxen}, which focused on the WEIRD perspective on user study participants in the HCI field; we focus on the UPS field.
Our study differs from the two closely related papers on UPS methodologies~\cite{TOCHI2021_Distler, arXiv2021_Kaur} in the following points\footnote{Note that this comparison is based on the simplified points and does not cover all the objectives of the papers.}:

\begin{itemize}
\setlength{\itemsep}{0pt}
\setlength{\parskip}{0pt}
\item We collected papers from more diverse conferences to gain a comprehensive understanding of UPS papers. 
\item We review more recent papers (2017--2021).
\item We analyze not only the degree to which the participants' countries were Western but also the degrees to which they were Educated, Industrialized, Rich, and Democratic. 
\item We unveil the degrees of Western-skew in research topics, conferences, and author affiliations.
\item We discuss the possible influence of participants' demographics on the results of UPS research and recommend future UPS research topics.
\end{itemize}

\noindent
Note that, recently, Mathew et al. reported work-in-progress results of author diversity in security and privacy papers, partly including UPS~\cite{WIPS2022_Mathew}.
Our study focuses on not only author diversity but also participant diversity.

\section{Methodology}

\begin{table*}[t] 
\caption{The number of UPS papers in each phase of our review process.}
\label{table:numberof_reviewed_paper}
\hbox to\hsize{\hfil
\footnotesize{
\begin{tabular}{l|r|r|r|r|r|r|r|r|r|r|r}\hline
Paper &  S\&P & SEC & CCS & NDSS & PETS & CHI & CSCW & SOUPS & EuroUSEC & USEC & Total \hspace{-0.7em} \\\hline\hline
\hspace{-0.7em} $\#$ Papers published for 5 years (Phase--1) & 426 & 701 & 751 & 403 & 313 & 3,471 & 1,268 & 143 & 57 & 54 & 7,587 \hspace{-0.7em} \\\hline
\hspace{-0.7em} $\#$ Potentially relevant papers (Phase--2) & 139 & 191 & 233 & 108 & 137 & 1,379 & 797 & 137 & 54 & 51 & 3,226 \hspace{-0.7em} \\\hline
\hspace{-0.7em} $\#$ Relevant papers (Phase--3) & 45 & 81 & 48 & 28 & 53 & 177 & 63 & 125 & 48 & 47 & 715 \hspace{-0.7em} \\\hline
\end{tabular}}\hfil}
\end{table*}

\begin{table}[t] 
\caption{The number of papers for detailed review by publication year.}
\label{table:numberof_reviewed_paper_peryear}
\hbox to\hsize{\hfil
\footnotesize{
\begin{tabular}{l|r|r|r|r|r|r}\hline
 & \multicolumn{1}{c|}{2017} & 
\multicolumn{1}{c|}{2018} & 
\multicolumn{1}{c|}{2019} & 
\multicolumn{1}{c|}{2020} & 
\multicolumn{1}{c|}{2021} & 
\multicolumn{1}{c}{Total}\\\hline
\multicolumn{1}{c|}{$\#$ Papers} & 117 & 128 & 119 & 150 & 201 & 715 \\\hline
\end{tabular}}\hfil}
\end{table}

\begin{table}[t] 
\caption{Extracted information for each paper.}
\label{table:extracted_information}
\hbox to\hsize{\hfil
\footnotesize{
\begin{tabular}{l|l}\hline
Category    & Item \\
\hline\hline
Publication & Title, conference, publication year \\\hline 
Participant & Number*, residence country*, education*, income*, \\
            & participant type (experts/non-experts)*\\
\hline
Author & Affiliation, affiliation country \\\hline
Method & Study method*, recruitment method*, research topic, \\
and Topic & design evaluation (y/n)*, attack feasibility evaluation (y/n)* \\\hline
\end{tabular}}\hfil}
\footnotesize{Asterisk marks mean the information that we extracted per user study rather than per paper. A single paper can conduct multiple user studies.}

\bigskip
\smallskip
\caption{Research methods and topics.}
\label{table:Options of method and topic items}
\hbox to\hsize{\hfil
\footnotesize{
\begin{tabular}{l|l}\hline
Item         & Classification  \\
\hline\hline
Study  & Survey, interview, lab study, online \\
method             & field experiment, trial deployment, other\\
\hline
Recruitment  & Institution, crowdsourcing, social media, city,\\
method       & mailing list/personal contacts, conference/event,\\
             & webpanel, source code repository, other, not mentioned\\
\hline
Research  & Overall, access control and privacy preference,\\
topic               & authentication, vulnerability/malware/incident response\\
               & privacy abuse, social engineering, privacy-enhancing\\
               & technologies, other\\
\hline
\end{tabular}}\hfil}
\end{table}

\subsection{Review Process}
\label{section:Review Process}

We followed the review process of recent representative review papers in the UPS field~\cite{TOCHI2021_Distler, SP2022_Obada}.
\medskip

\noindent\textbf{Criteria.}
Since we are interested in geographic diversity of participants recruited in UPS papers, we decided to review papers that meet three criteria: full-length papers (i) that addressed issues of human factors and computer security/privacy as the main research objectives, (ii) that recruited voluntary participants, and (iii) that reported the total number of participants.
We removed the following papers because they did not meet criterion--(ii):
\begin{itemize}
\setlength{\itemsep}{0pt}
\setlength{\parskip}{0pt}
\item The papers that analyzed online public data, such as online services design and social media posts.
\item The papers that analyzed behavioral logs of employees (e.g., response to simulated phishing) or that of actual service users (e.g., response to  warnings displayed on browser).
\item The papers in which the authors conducted ethnographic observations.
\end{itemize}

\noindent\textbf{Phase--1: Conference selection.}
We decided to review all conferences that were selected in either or both of the two recent representative papers that comprehensively reviewed UPS papers~\cite{TOCHI2021_Distler, SP2022_Warford}: S\&P, USENIX Security (SEC), CCS, NDSS, and PETS (top and second tier conferences in security and privacy), CHI and CSCW (top-tier conferences in HCI), and SOUPS, EuroUSEC, and USEC (a conference and workshops that focus on UPS).
The total number of full-length papers of the above 10 conferences between 2017--2021 was 7,587, as shown in Table~\ref{table:numberof_reviewed_paper}.

\noindent\textbf{Phase--2: Screening for potentially relevant papers.}
The dataset of Phase--1 included numerous non-UPS security papers (e.g., cryptography theory) and non-UPS HCI papers. 
To exclude the papers that did not meet the criteria, we filtered papers with the following search queries: \textit{(security OR privacy) AND (recruit* OR participant* OR respondent*)}. 
Since we were especially interested in the UPS papers that recruited participants, we used the terms \textit{recruit}, \textit{participant}, and \textit{respondent} instead of the terms relevant to all UPS papers such as \textit{user} or \textit{usability}. 
Whereas a prior study~\cite{TOCHI2021_Distler} conducted a similar search for titles or abstracts, we conducted it for full texts to prevent inappropriately excessive exclusion. 
We explain the detailed procedure in the Appendix~\ref{section:appendix_screening_detailed}. 
Our search query resulted in 3,226 potentially relevant papers.

\noindent\textbf{Phase--3: Identifying relevant papers.} 
The first author reviewed the titles, abstracts, and body text of the 3,226 papers and removed papers that did not meet our criteria. 
Every time the first author found a paper where it was unclear whether it met the criteria, the authors discussed and resolved the issue. 
The last author independently double-coded 323 randomly selected papers (10\%). 
The inter-rater reliability was Cohen's kappa $k=$ .86, which is considered to be an excellent level of agreement~\cite{IRR}. 
After this manual screening, we finally identified 715 relevant papers\footnote{The full list of papers is available in the Appendix~\ref{section:appendix_papers_list}.}. 
Table~\ref{table:numberof_reviewed_paper_peryear} lists the numbers of identified UPS papers in each publication year. 
The number of UPS papers noticeably increased in 2020--2021, which Distler et al. and Kaur et al. do not cover~\cite{TOCHI2021_Distler, arXiv2021_Kaur}.

\noindent\textbf{Phase--4: Detailed review.}
The first author reviewed each paper in detail and extracted information about the publication, participant, author, and method/topic listed in Table~\ref{table:extracted_information}.
Note that we did not extract information of pilot studies.
The coding items were based on the Linxen study~\cite{CHI2021_Linxen}, and some of them were newly added in this work for UPS papers. 
The newly added items are described below.

\begin{itemize}
\setlength{\itemsep}{3pt}
\setlength{\parskip}{0pt}
\item 
\textbf{Participant type} is whether the participants are experts or non-experts (end users)~\cite{arXiv2021_Kaur}.
\item 
\textbf{Study method} and \textbf{recruitment method} consist of the classification listed in Table~\ref{table:Options of method and topic items}.
We created this classification with reference to the existing classification~\cite{TOCHI2021_Distler, arXiv2021_Kaur}.
\item 
\textbf{Research topics} is the classification (seven distinct topics and ``other'') listed in Table~\ref{table:Options of method and topic items}. 
We improved an existing classification~\cite{TOCHI2021_Distler} by focusing on technology (details are described in the Appendix~\ref{section:appendix_codingrule_topic}). 
\item 
\textbf{Design evaluation} is whether the authors experimented with technology or service design. 
For example, papers that compared multiple interfaces or framing with a between-group or within-group approach, or papers that clearly stated that usability evaluations have been performed are coded ``yes.''
\item 
\textbf{Attack feasibility evaluation} is unique to UPS: whether the authors conducted a user study for only demonstrating the feasibility of their proposed attack and did not ask participants about their perceptions of security and privacy issues.
\end{itemize}

\noindent
We extracted information per user study rather than per paper if a single paper recruited participants separately for multiple user studies with different study methods (e.g., a survey and interview) or with different purposes (e.g., one survey to identify user concerns and another to evaluate the proposed UI), as shown in Table~\ref{table:extracted_information}.

Some papers were ambiguous in their description of the recruitment process and participant information.
In addition, it is potentially difficult to classify research topics of each paper.
To reduce coders' subjectivity, the first and last authors discussed many times to develop a codebook and resolve the discrepancies in their interpretations.
The last author independently double-coded 72 randomly selected papers (10\%). The Cohen's kappa of our review was $k$ = .85--.97, which is considered to be an excellent level of agreement~\cite{IRR}.

\subsection{Analysis}
\label{section:method_analysis}

We followed the analysis framework of the WEIRD acronym used in the Linxen study~\cite{CHI2021_Linxen}.
We used the same year's national statistics as the Linxen study to directly compare the results of the Linxen study (i.e., the HCI field) with our results (i.e., the UPS field)\footnote{The statistical data was provided from the authors of the Linxen study.}.

\smallskip
\noindent\textbf{Participant sample.} 
The number of participant samples refers to the number of countries from which there were participants who were recruited for a user study. 
A single paper can report multiple participant samples, i.e., a single paper can recruit participants from multiple countries.

\smallskip
\noindent\textbf{WEIRD.} The WEIRD acronym (Western, Educated, Industrialized, Rich, and Democratic) is the set of indicators for measuring the diversity of participants' countries. 
Just because a country is non-Western does not necessarily mean it is non-EIRD, i.e., there are certain non-W and EIRD countries such as Israel, Japan, and South Korea.
\begin{itemize} 
\setlength{\itemsep}{5pt}
\setlength{\parskip}{0pt}
\item \textbf{Western.}
We classified participants' countries described in each paper into Western or non-Western using Huntington's classification~\cite{Huntington}.
This classification is based on civilizations in the post-Cold War era, and Western is centered in Europe and North America.  
Note that while we adopt this classification for our comparison, it has been criticized in the political science community~\cite{criticism1, criticism2} and does not always agree with the general perception of Western.
\begin{itemize} 
\setlength{\itemsep}{0pt}
\setlength{\parskip}{0pt}
\item Western (examples): U.S., Canada, UK, Germany, Netherlands, Switzerland, Sweden, and Australia.
\item Non-Western (examples): India, China, Japan, South Korea, Israel, Saudi Arabia, Russia, Pakistan, Bangladesh, Mexico, Brazil, and South Africa.
\end{itemize} 
\item \textbf{Educated.}
We comprehensively explored participants' education levels by analyzing two indicators:
(a) the mean years of schooling per country~\cite{education_stats} and (b) specific educational levels of each participant if the information was provided in a paper.
\item \textbf{Industrialized.}
We used GDP (gross domestic product) per country, which is regarded as the most representative indicator for assessing the development and progress of a national economy~\cite{GDP2}.
To adjust for differences in purchasing power among countries, we used GDP based on PPP (purchasing power parity)~\cite{GDP1}.
\item \textbf{Rich.}
We comprehensively explored participants' wealth by analyzing two indicators:
(a) the country's GNI (gross national income) based on PPP~\cite{GNI1} and (b)
specific income levels of each participant if the information was provided in a paper.
\item \textbf{Democratic.}
We used the political rights rating, which indicates countries' degrees of democracy and covers electoral process, political pluralism and participation, and functioning of government~\cite{politicalrights}.
\end{itemize}

\smallskip
\noindent\textbf{Normalization by country population.}
To address which countries are over- or under-represented in UPS papers relative to population, we normalized the number of participant samples and participants by the country's population~\cite{worldpopulation}.
Specifically, we calculated the normalized ratio of participant samples ({\small$\psi_{\scalebox{0.9}[1.0]{\textit{s}}}$}) and that of participants ({\small$\psi_{\scalebox{0.9}[1.0]{\textit{p}}}$}) as follows:
{\small $$\psi_{\scalebox{0.9}[1.0]{\textit{s}}} = \frac{n_{\scalebox{0.9}[1.0]{\textit{samples}}}\hspace{0.2em}\textrm{(country)} \hspace{0.5em}\cdot\hspace{0.5em} \textrm{population (world)} }{n_{\scalebox{0.9}[1.0]{\textit{samples}}}\hspace{0.2em}\textrm{(total)} \hspace{0.5em}\cdot\hspace{0.5em} \textrm{population (country)}},$$}
{\small $$\psi_{\scalebox{0.9}[1.0]{\textit{p}}} = \frac{n_{\scalebox{0.9}[1.0]{\textit{participants}}}\hspace{0.2em}\textrm{(country)} \hspace{0.5em}\cdot\hspace{0.5em} \textrm{population (world)} }{n_{\scalebox{0.9}[1.0]{\textit{participants}}}\hspace{0.2em}\textrm{(total)} \hspace{0.5em}\cdot\hspace{0.5em} \textrm{population (country)}}$$}
\noindent
The values of {\small$\psi_{\scalebox{0.9}[1.0]{\textit{s}}}$} and {\small$\psi_{\scalebox{0.9}[1.0]{\textit{p}}}$} are at least 0. 
A value of {\small$\psi_{\scalebox{0.9}[1.0]{\textit{s}}}<1$} or {\small$\psi_{\scalebox{0.9}[1.0]{\textit{p}}}<1$} indicates the country is under-represented, while a value of {\small$\psi_{\scalebox{0.9}[1.0]{\textit{s}}}>1$} or {\small$\psi_{\scalebox{0.9}[1.0]{\textit{p}}}>1$} indicates that the country is over-represented corresponding to the country's population as a percentage of the world's population.
For example, {\small$\psi_{\scalebox{0.9}[1.0]{\textit{s}}}=2$} indicates that the country was over-studied twice as much relative to its population size.

\smallskip
\noindent\textbf{Correlation analysis.}
To understand to what extent participant samples tend to be from EIRD countries, we calculated the Kendall's tau rank correlations between {\small$\psi_{\scalebox{0.9}[1.0]{\textit{s}}}$} and each indicator of EIRD (e.g., mean schooling years and GDP).
For instance, if there is a significantly positive correlation between {\small$\psi_{\scalebox{0.9}[1.0]{\textit{s}}}$} and the mean schooling years, we conclude that participant samples of UPS papers come from generally educated countries.

\smallskip
\noindent \textbf{Author diversity.}
We identified the country of each author's affiliation on the basis of their affiliated institution listed first. Each paper was classified either as Western-affiliated authors only, both Western-affiliated and non-Western affiliated authors, or non-Western affiliated authors only.

\smallskip
\noindent
\textbf{Differences among conferences and research topics.}
To reveal the characteristics of conferences, we compared the percentage of papers that were published only by Western-affiliated authors, papers that recruited only Western participant samples and papers that reported the participants' countries across conferences. 
Similarly, we compared those percentages across research topics, participant type, design evaluation (y/n), and attack feasibility evaluation (y/n).

\section{Results}
\label{section:results}

\subsection{Western}
\label{section:results_western}

Of our dataset, 51.89\% (371/715) papers explicitly reported the participants' countries.
In 15.94\% (114/715) of papers, although the authors did not mention the participants' countries, we were able to infer them from the authors' affiliations (e.g., if authors mentioned that participants were recruited locally and all the authors were at institutions in the same country). 
For the remaining 32.17\% (230/715) of papers, we were not able to confidently identify the participants' countries because the authors provided no or unclear descriptions. 
For example, there were papers (i) that described only that participants were recruited locally or a user study was conducted in person, but not all the authors were in the same country; (ii) that described only that participants were recruited thorough social media, crowdsourcing, or personal contacts (i.e., recruitment methods that may include people outside of the authors' country, especially for online studies); and (iii) that described only that participants were recruited at a security conference held in a specific country (some attendees may be from outside of the country).
We provide a flowchart for coding in participants' countries in Appendix~\ref{section:appendix_codingrule_country}.

\begin{savenotes}
\begin{table}[t] 
\caption{The number of participant samples and participants.}
\label{table:results_overview}
\hbox to\hsize{\hfil
\footnotesize{
\begin{tabular}{l|r|r|r|r}\hline
\multirow{2}{*}{Countries} & \multicolumn{2}{c|}{$n_{\scalebox{0.9}[1.0]{\textit{samples}}}$} & \multicolumn{2}{c}{$n_{\scalebox{0.9}[1.0]{\textit{participants}}}$} \\\cline{2-5}
 & \multicolumn{1}{c|}{Ours} & \multicolumn{1}{c|}{\cite{CHI2021_Linxen}} & \multicolumn{1}{c|}{Ours} & \multicolumn{1}{c}{\cite{CHI2021_Linxen}\footnote{This value was not mentioned in the Linxen study, but we compiled it using the artifact data published by the authors.}}\\\hline \hline
Western & 624 & 1,102 & 183,828 & 174,041 \\
non-Western & 167 & 405 & 38,208 & 75,068 \\
Reported as ``other'' & -- & -- & 4,836 & n/a \\\hline
\% Western & 78.89\% & 73.13\% & 82.79\% & 69.87\% \\\hline
\end{tabular}}\hfil}
\end{table}
\end{savenotes}

Table~\ref{table:results_overview} shows the number of participant samples and participants among 485 papers for which we were able to identify or infer the participants' countries.
In terms of the number of participant samples, a large majority of UPS papers involved Western participant samples (78.89\%, 624/791). 
This is higher than the results in the Linxen study~\cite{CHI2021_Linxen} that reviewed studies in the HCI field (73.13\%). 
In terms of the number of participants, 226,872 were involved in UPS user studies. 
Except for 4,836 participants who were reported as ``and other'' and whose details were not available, 82.79\% (183,828/222,036) of participants were from Western countries. 
Although our analysis methods are the same as those in the Linxen study~\cite{CHI2021_Linxen}, please note that this is a conservative estimate, given that we only included papers for which we were able to identify or infer participants' countries. 
If researchers recruited non-Westerners, they would probably have reported it because it is an uncommon practice.
Thus, we are concerned that the true percentage of the Western participant samples is much higher.

\begin{figure}[!t]
\begin{center}
\includegraphics[width=0.9\linewidth]{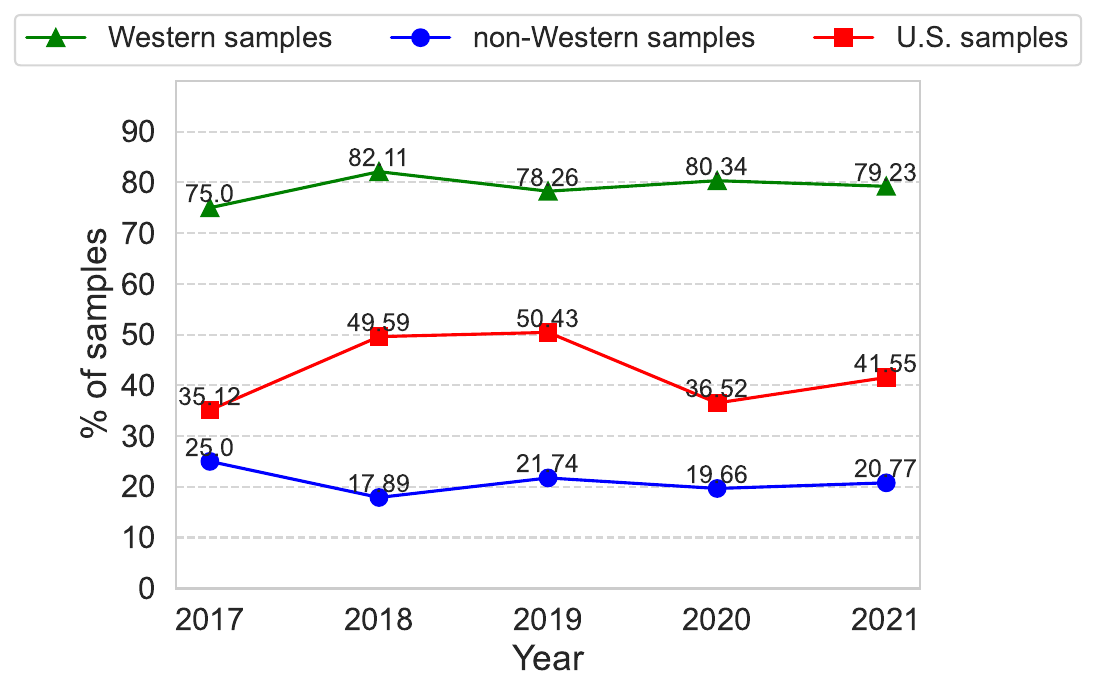}
\end{center}
\caption{Temporal changes in participant samples.}
\label{Figs:fiveyear}
\end{figure}

Figure~\ref{Figs:fiveyear} shows the temporal changes in the percentages of Western, non-Western, and U.S. participant samples over the five-year period.
Whereas the Linxen study reported that the percentage of non-Western participant samples in CHI papers largely increased (16.31\% to 30.24\%, 1.85x) between 2016--2020, our results unveiled that the percentage in UPS papers between 2017--2021 slightly decreased (25.00\% to 20.77\%, 0.83x).

\begin{savenotes}
\begin{table*}[t] 
\caption{Top five countries by the number of participant samples (left), the normalized ratio of participant samples (middle), and the number of participants (right).}
\label{table:results_ranking}
\hbox to\hsize{\hfil
\footnotesize{
\begin{tabular}{lrrr|lrrr|lrrr}\hline
\multicolumn{4}{c|}{Top countries by $n_{\scalebox{0.9}[1.0]{\textit{samples}}}$} & \multicolumn{4}{c|}{Top countries by $\psi_{\scalebox{0.9}[1.0]{\textit{s}}}$} & \multicolumn{4}{c}{Top countries by $n_{\scalebox{0.9}[1.0]{\textit{participants}}}$}  \\\hline
\multicolumn{1}{c}{Country} & \multicolumn{1}{c}{$n_{\scalebox{0.9}[1.0]{\textit{samples}}}$} & \multicolumn{1}{c}{\%} & \multicolumn{1}{c|}{$\psi_{\scalebox{0.9}[1.0]{\textit{s}}}$} & \multicolumn{1}{c}{Country} & \multicolumn{1}{c}{$n_{\scalebox{0.9}[1.0]{\textit{samples}}}$} & \multicolumn{1}{c}{\%} & \multicolumn{1}{c|}{$\psi_{\scalebox{0.9}[1.0]{\textit{s}}}$} & \multicolumn{1}{c}{Country} & \multicolumn{1}{c}{$n_{\scalebox{0.9}[1.0]{\textit{participants}}}$} & \multicolumn{1}{c}{\%} & \multicolumn{1}{c}{$\psi_{\scalebox{0.9}[1.0]{\textit{p}}}$} \\\hline\hline
U.S. & 329 & 41.59 & 9.79 & Andorra & 1 & 0.13 & 127.98 & U.S. & 142,908 & 62.99 & 14.83\\
Germany & 76 & 9.61 & 8.94 & Cyprus & 2 & 0.25 & 16.33 & Japan\footnote{The number of Japanese participant samples was small ($N$=6), but one user study recruited a very large number of participants ($N$=20,645).} & 21,979 & 9.69 & 5.97 \\
UK & 67 & 8.47 & 9.73 & Luxembourg & 1 & 0.13 & 15.77 & Germany & 21,857 & 9.63 & 8.96\\
Canada & 33 & 4.17 & 8.61 & Austria & 9 & 1.14 & 9.85 & UK & 9,297 & 4.10 & 4.71\\
India & 26 & 3.29 & 0.19 & U.S. & 329 & 41.59 & 9.79 & China & 4,994 & 2.20 & 0.12\\\hline
\end{tabular}}\hfil}
\end{table*}
\end{savenotes}

\begin{figure}[!t]
\begin{center}
\includegraphics[width=\linewidth]{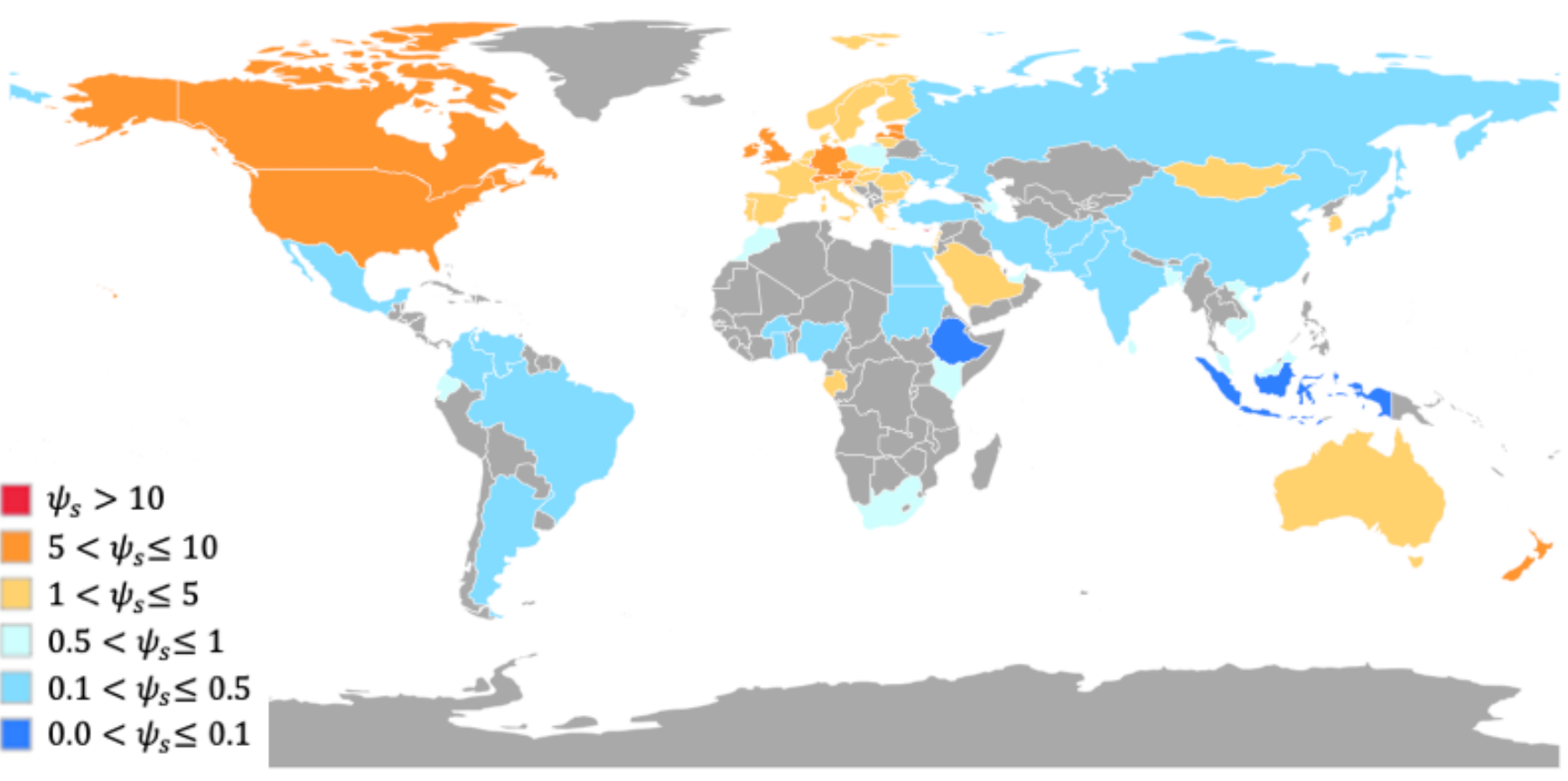}
\end{center}
\caption{Distribution of normalized participant samples.}
\label{Figs:map}
\end{figure}

Table~\ref{table:results_ranking} shows the top five countries by the number of participant samples, normalized ratio of participant samples by country population, and the number of participants, which enabled us to grasp over- and under-represented countries. 
Participant samples of UPS papers heavily skewed toward four Western countries: U.S. has been the most frequently studied, followed by Germany, United Kingdom, and Canada.
Although these top four countries are the same as those for the CHI papers (the Linxen study), their shares are higher in the UPS papers (63.84\%) than in the CHI papers (55.61\%). 
Considering the U.S. population as a percentage of the world population, U.S. participant samples in the UPS field are over-represented by 9.79 times. 
Figure~\ref{Figs:map} shows the worldwide distribution of a normalized ratio of participant samples. 
Participants in the UPS papers between 2017 and 2021 were from only 79 countries, and many countries are completely missing from this map, especially African ones. 
South America, the Middle East, and Asia also have many countries that are missing or under-represented.

In the following two subsections, indicators of E, I, R, and D are analyzed independently. 
The percentage of participant samples that are from fully WEIRD (i.e., using AND as the logical operator) countries is presented in the Appendix~\ref{section:appendix_WEIRD_with_AND_Operators}.

\subsection{Educated}
\label{section:results_educated}

\begin{table}[t] 
\caption{Correlations of the normalized ratio of participant samples $\psi_{\scalebox{0.9}[1.0]{\textit{s}}}$ with each indicator of Educated, Industrialized, Rich, and Democratic.}
\label{table:results_correlation}
\hbox to\hsize{\hfil
\footnotesize{
\begin{tabular}{l|r|r}\hline
\multicolumn{1}{c|}{Indicators} & \multicolumn{1}{c|}{$r$} & \multicolumn{1}{c}{$r$~\cite{CHI2021_Linxen}} \\\hline\hline
Educated (mean schooling year) & .57*** & .46***\\\hline
Industrialized (GDP) & .49***  & .50***\\\hline 
Rich (GNI) & .52*** & .50***\\\hline
Democratic (political rights) & .44*** & .50*** \\\hline 
\end{tabular}}\hfil}
\footnotesize{Significance levels: *$p <$ .05, **$p <$ .01, ***$p <$ .001.}
\end{table}

Table~\ref{table:results_correlation} indicates the correlation between countries' normalized ratio of participant samples and countries' mean years of schooling (see Analysis of Education-(a) in Section~\ref{section:method_analysis}). 
We found a positive correlation ($r=$.57, $p<$.001), i.e., most participant samples of UPS papers come from countries with generally highly educated populations.

Next, we analyzed the participants' education levels on the basis of the detailed descriptions in each paper (see Analysis of Education-(b) in Section~\ref{section:method_analysis}). 
In our dataset, only 34.69\% (248/715) of papers explicitly reported their participants' education levels.
Although 14.13\% (101/715) of papers did not report their participants' education levels, we were able to infer the educational levels of participants because the authors mentioned that they recruited university students. 
In the remaining 51.19\% (366/715) of papers, we were not able to find descriptions of their participants' education levels.
Among the participants recruited in the 248 papers in which the authors explicitly reported their participants' education levels, 71.38\% (79,539/111,436) of participants had a college-level or higher education, 20.89\% (23,282/111,436) had a secondary or lower education, and the remaining 7.73\% (8,615/111,436) were reported as ``other'' or ``prefer not to answer.''
The participants of UPS papers (71.38\% of the participants were college-educated) are as highly educated as the participants in CHI papers (69.93\% of the participants were college-educated). 
We believe that the skew toward highly educated participants is related to the recruiting methods (described in Section~\ref{section:results_author}): (1) many papers recruited participants within the authors' institutions including universities, (2) many papers recruited participants through crowdsourcing such as MTurk (e.g., MTurk workers tend to be more educated than the general U.S. population~\cite{SP2019_Redmiles}), and (3) there is a certain percentage (14.96\%) of papers that recruited experts (who are generally highly educated).

\subsection{Industrialized, Rich, and Democratic}
\label{section:results_industrialized_and_rich}

Table~\ref{table:results_correlation} shows the results of the correlations between countries' normalized ratios of participant samples and countries' GDP and GNI (see Analysis of Rich-(a) in Section~\ref{section:method_analysis}).
We found positive correlations for both GDP ($r=$.49, $p<$.001) and GNI ($r=$.52, $p<$.001), i.e., most participant samples of UPS papers come from industrialized and rich countries. 

Next, we analyzed the participants' income on the basis of the detailed descriptions in each paper (see Analysis of Rich-(b) in Section~\ref{section:method_analysis}). 
However, only 5.59\% (40/715) reported the income levels of their participants. 
As in the Linxen study, we are also not able to directly establish whether the wealth of participants is representative of the general population.

Finally, as shown in Table~\ref{table:results_correlation}, we found a positive correlation between countries' normalized ratio of participant samples and countries' political rights ($r=$.44, $p<$.001), i.e., most participant samples of UPS papers come from countries with a high degree of freedom in terms of political rights.

\begin{table*}[t] 
\caption{Author affiliations and study/recruitment method.}
\label{table:results_authors}
\hbox to\hsize{\hfil
\footnotesize{
\begin{tabular}{l;{0.4pt/1pt}l;{0.4pt/1pt}r;{0.4pt/1pt}r;{0.4pt/1pt}l;{0.4pt/1pt}l}\hline
\multirow{2}{*}{Authors} & \multirow{2}{*}{Participants} & \multicolumn{2}{c;{0.4pt/1pt}}{Study Method} & \multicolumn{2}{c}{Recruitment Method} \\\cdashline{3-6}[0.4pt/1pt]
 & & \multicolumn{1}{c;{0.4pt/1pt}}{Survey} & \multicolumn{1}{c;{0.4pt/1pt}}{Int/Lab} & \multicolumn{1}{c;{0.4pt/1pt}}{Top1} & \multicolumn{1}{c}{Top2} \\\hline \hline
\multicolumn{1}{|c|}{\multirow{3}{5.2em}{Western only}} & \multicolumn{1}{|l|}{Only authors' countries (72.65\%)} & 35.73\% & \multicolumn{1}{r|}{55.88\%}  & Institution (32.85\%) & \multicolumn{1}{l|}{CS (27.58\%)} \\\cline{3-6}
\multicolumn{1}{|c|}{} & \multicolumn{1}{|l|}{Other countries: Western only (7.84\%)} & 57.78\% & \multicolumn{1}{r|}{37.78\%} & CS (40.00\%) & \multicolumn{1}{l|}{Social media (28.89\%)} \\\cline{3-6}
\multicolumn{1}{|c|}{} & \multicolumn{1}{|l|}{Other countries: non-Western included (7.49\%)} & 60.47\% & \multicolumn{1}{r|}{32.56\%}  & Social media (32.56\%) & \multicolumn{1}{l|}{CS (30.23\%)} \\\cdashline{1-2}[0.4pt/1pt] \cline{3-6}
\multicolumn{1}{|c|}{\multirow{4}{5.2em}{non-Western\\included}} & \multicolumn{1}{|l|}{Only authors' countries: Western only (2.79\%)} & 62.50\% & \multicolumn{1}{r|}{31.25\%} & CS (62.50\%) & \multicolumn{1}{l|}{Institution (12.50\%)} \\\cline{3-6}
\multicolumn{1}{|c|}{} & \multicolumn{1}{|l|}{Only authors' countries: non-Western included (5.40\%)} & 25.81\% & \multicolumn{1}{r|}{58.06\%} & Institution (51.61\%) & \multicolumn{1}{l|}{ML/PC (22.58\%)} \\\cline{3-6}
\multicolumn{1}{|c|}{} & \multicolumn{1}{|l|}{Other countries: Western only (1.22\%)} & 85.71\% & \multicolumn{1}{r|}{14.29\%} & CS (85.71\%) & \multicolumn{1}{l|}{Institution (14.29\%)} \\\cline{3-6}
\multicolumn{1}{|c|}{} & \multicolumn{1}{|l|}{Other countries: non-Western included (2.61\%)} & 33.33\% & \multicolumn{1}{r|}{53.33\%} & \multicolumn{2}{l|}{City (26.67\%), ML/PC (26.67\%)} \\\hline
\end{tabular}}\hfil}
\footnotesize{We show the percentages of user studies rather than that of papers because multiple user studies (e.g., interview and survey) can be conducted in a single paper. 
We focused on the user studies in which we were able to identify or infer the participants' countries.
Figures represent the percentages of user studies within the sections separated by solid lines. 
``Int/Lab'': interview or lab study. ``CS'': crowdsourcing. ``ML/PC'': mailing list or personal contacts.}
\end{table*}

\subsection{Author Diversity and Methodology}
\label{section:results_author}

Of our dataset, 86.57\% (619/715) of the papers consisted exclusively of authors affiliated with institutions in Western countries, which is slightly even more Western-skewed than the Linxen study (83.08\%\footnote{This value was not mentioned in the Linxen study, but we compiled it using the artifact data published by the authors.}). 
9.93\% (71/715) of the papers consisted of both Western-affiliated and non-Western affiliated authors, and the remaining 3.50\% (25/715) of the papers consisted of only non-Western affiliated authors. 
The percentage of papers that included authors affiliated with institutions in non-Western countries remained roughly the same: from 14.53\% (17/117) in 2017 to 13.43\% (27/201) in 2021.

Table~\ref{table:results_authors} shows the relationships of authors' affiliation countries, participants' countries, study methods, and recruitment methods. 
In 80.84\% (464/574) of user studies, the authors recruited participants from only the countries of their affiliations. 
The possible reasons that participants samples are skewed towards Western countries as we presented in Section~\ref{section:results_western} include: (1) the countries of authors' affiliations in UPS papers are skewed toward Western ones and (2) authors tend to recruit participants in the countries of their affiliations.

In most user studies, surveys, interviews, or lab studies are conducted. 
As shown in Table~\ref{table:results_authors}, when authors recruited participants from other countries, they were more likely to conduct surveys than lab studies or interviews. 
Lab studies and interviews require researchers to communicate immediately/interactively with participants, and thus linguistic barriers are a serious problem. 
Moreover, geographic barriers also affect the choice of study method. 
While lab studies and interviews essentially require participants to physically come to a specific location, surveys are conducted online in most cases. 
Note that interviews can also be conducted online now thanks to video conferencing services.

For recruiting participants in the authors' country, authors tend to recruit within their institutions (which is a university in most user studies). 
We found crowdsourcing platforms are commonly used for recruiting participants both inside and outside of authors' countries. 
However, registered workers of commonly used crowdsourcing platforms such as Amazon Mechanical Turk (MTurk) and Prolific are skewed toward Westerners.
Specifically, MTurk workers are mostly from the U.S., followed by India, whereas Prolific workers are primarily from the UK, followed by the U.S. and other Europe countries~\cite{workercountries1, workercountries2}. 
Thus, such popular platforms contribute little to the recruitment of non-Westerners.
Specifically, the Indian participant sample was the fifth largest in our dataset thanks to India's high numbers of MTurkers and English speakers as shown in Table~\ref{table:results_ranking}, but non-Western countries except India are less likely to be studied by Western authors. 
We were interested in how Western-affiliated authors were reaching non-Western participants (excluding Indians, who are easily accessible via MTurk). 
In such user studies, the authors advertised their user study on social media, or one of the authors, who belonged to a Western affiliation but had a non-Western background, was in charge of recruitment in the city or through the author's social network~\cite{CHI2017_Ahmed, CSCW2017_Ahmed}.

\subsection{Differences among Conferences}
\label{section:results_differences_by_venues}

\begin{table}[t] 
\caption{Characteristics of conferences.}
\label{table:results_conference}
\hbox to\hsize{\hfil
\footnotesize{
\begin{tabular}{l|r|r|r|r}\hline
\multicolumn{1}{c|}{\multirow{2}{*}{Conference}} & \multicolumn{1}{c|}{\% Country} & \multicolumn{1}{c|}{\% Attack} & \multicolumn{1}{c|}{\% W-only} & \multicolumn{1}{c}{\% W-only} \\
 & \multicolumn{1}{c|}{Reported} & \multicolumn{1}{c|}{Feasibility} & \multicolumn{1}{c|}{Authors} & \multicolumn{1}{c}{Samples} \\\hline\hline
\hspace{-0.6em} S\&P field & 56.08 (36.86) & 18.82 & 81.96 & 85.31 \\\hline
\hspace{-0.6em} HCI field & 75.83 (61.25) & 0.42 & 89.58 & 84.62 \\\hline
\hspace{-0.6em} UPS-focused &  72.73 (59.09) & 0.45 & 88.64 & 86.25 \\\hline
\end{tabular}}\hfil}
\footnotesize{``S\&P field'': S\&P, SEC, CCS, NDSS, and PETS. ``HCI field'': CHI and CSCW. ``UPS-focused'': SOUPS, EuroUSEC, and USEC. ``Country Reported'' column: the percentage of the papers in which we were able to identify or infer the participants' countries (the figures in parentheses indicate the papers in which we were able to identify them).}
\end{table}

Table~\ref{table:results_conference} shows the characteristics of authors and participant samples by conferences. 
Both authors' affiliations and participants were skewed toward Westerners at every conference.  
The percentages of the papers in which the authors explicitly reported their participants' countries are lower in S\&P, SEC, CCS, NDSS, and PETS than in HCI conferences (CHI and CSCW) and UPS-focused conferences (SOUPS, EuroUSEC, and USEC). 
Those five conferences (S\&P, SEC, CCS, NDSS, and PETS) have relatively higher percentages of the papers in which the authors recruited participants only for demonstrating the feasibility of their proposed attack.

\subsection{Research Topics and Methodology}
\label{section:Differences by Research Topics}

\begin{table}[t] 
\caption{Differences among participant type and study objectives.}
\label{table:results_topic1}
\hbox to\hsize{\hfil
\footnotesize{
\begin{tabular}{l|r|r|r}\hline
\multirow{2}{*}{} & \multicolumn{1}{c}{\% Studies} & \multicolumn{1}{|c|}{\% Country} & \multicolumn{1}{|c}{\% W-only} \\
 & \multicolumn{1}{c}{} & \multicolumn{1}{|c|}{Reported} & \multicolumn{1}{|c}{Samples} \\\hline \hline
Participants -- Non-experts & 85.04 & 70.67 (53.07) & 85.38 \\
Participants -- Experts & 14.96 & 53.97 (38.10) & 77.94 \\\hline
Design evaluation -- Yes & 39.55 & 68.47 (41.44) & 83.78 \\
Design evaluation -- No & 60.45 & 67.98 (56.97) & 84.97 \\\hline
Attack feasibility -- Yes & 7.01 & 44.07 (15.25) & 92.31 \\
Attack feasibility -- No & 92.99 & 69.99 (53.51) & 84.12\\\hline
\end{tabular}}\hfil}
\footnotesize{We show the percentages of user studies rather than those of papers because authors can conduct multiple user studies in a single paper (e.g., researchers recruit non-experts for the first study and experts for the second study).}
\end{table}

\begin{table}[t] 
\caption{Differences among research topics.}
\label{table:results_topic2}
\hbox to\hsize{\hfil
\footnotesize{
\begin{tabular}{p{2.7cm}|r|r|r}\hline
\multirow{2}{*}{Topic} & \multicolumn{1}{c}{\% Papers} & \multicolumn{1}{|c|}{\% Country} & \multicolumn{1}{|c}{\% W-only} \\
 & \multicolumn{1}{c}{} & \multicolumn{1}{|c|}{Reported} &\multicolumn{1}{|c}{Samples} \\\hline \hline
Access control and privacy preference & 31.47 & 75.11 (63.56) & 90.53  \\\hline
Authentication & 17.34 & 65.32 (33.06) & 82.72 \\\hline
Vulnerability, malware, and incident response & 13.43 & 56.25 (41.67) & 77.78  \\\hline
Overall & 11.33 & 77.78 (67.90) & 85.71 \\\hline
Privacy-enhancing technologies & 9.79 & 60.00 (45.71) & 85.71 \\\hline
Social engineering & 6.29 & 75.56 (64.44) & 79.41 \\\hline
Privacy abuse & 5.45 & 76.92 (66.67) & 80.00 \\\hline
Other & 4.90 & 34.29 (14.29) & 91.67\\\hline
\end{tabular}}\hfil}
\footnotesize{We show the percentages of papers rather than those of user studies because research topics in a single paper can be classified into one code, even if the paper conducted multiple user studies.}
\end{table}

Table~\ref{table:results_topic1} shows the percentages of user studies that recruited only Western samples by participant type and study objectives. 
In terms of participant type (i.e., non-experts or experts), we found that the percentages of user studies that recruited only Western samples are higher among the user studies that focused on non-experts than among the user studies that focused on experts. 
This result was influenced by the fact that (1) more than 40\% (40.48\%) of user studies that focused on experts recruited ``developers,'' and (2) for recruiting developers, Western-affiliated authors succeeded in accessing non-Western developers by using Freelancer.com (on which many non-Western developers are registered) and developer contacts published on source code repositories (e.g., GitHub), and app stores (e.g., Google's Play Store). 
However, if the user studies that recruited developers are excluded, the participants of the remaining user studies (e.g., user studies that recruited system/network administrators and security experts) skewed heavily towards Western participant samples (88.57\%).

When comparing the user studies that conducted design evaluation (e.g., evaluated usability of the proposed prototype or effectiveness of the proposed nudges for security) with the user studies that did not conduct design evaluation, we found no difference in the degree of skewness toward Western participant samples.

The user studies conducted only for demonstrating the feasibility of the proposed attack account for 7.01\% (59/842). 
Regarding such user studies, we found that (1) authors were less likely to report participants' countries explicitly (15.25\%), and (2) authors were likely to recruit only Western samples (92.31\%). 
The heavy Western skew of participant samples among attack-feasibility studies can be explained by the following fact: MTurk and lab studies, which are prone to demographic skew (e.g., the workers in MTurk are mostly from the U.S., and university researchers in Western countries are likely to recruit students at their universities for a lab study), were often used for demonstrating the feasibility of attacks.
We further analyzed the studies of attack feasibility (56 papers including 59 user studies) and confirmed that attack feasibility can be influenced by participant demographics in the majority of studies (54.24\%, 32/59).
Two types of attack made up the majority: (i) targeting automatic speech recognition (15.25\%, 9/59) and (ii) typing/keystroke inference (13.56\%, 8/59). 
Because most such studies implicitly leveraged English audio or input in English, the attack feasibility may vary depending on the language used by the participants.
For example, participants with low English proficiency may have difficulty recognizing malicious English audio. 
Additionally, for example, the keyboard layout and character input/conversion method used in each linguistic area can affect the feasibility of typing/keystroke inference. 
Therefore, authors who demonstrate attack feasibility should also properly report participants' demographic information such as their country of residence and language proficiency.

Table~\ref{table:results_topic2} shows the percentages of papers that recruited only Western participant samples by research topics. 
In UPS, ``access control and privacy preference'' was most frequently studied, followed by ``authentication,'' ``vulnerability, malware, and incident response,'' and ``overall.'' 
The Western skew is pervasive in every topic. 
The degree of Western skew is relatively higher in ``access control and privacy preference'' and ``other.''
On the other hand, in three research topics, the degree of Western skew is relatively lower: (i) for ``vulnerability, malware, and incident response,'' researchers often recruited developers, and as mentioned above, researchers succeeded in accessing non-Western developers; (ii) regarding ``social engineering'', several researchers realized the importance of investigating the phishing susceptibility of non-Westerners (e.g., emerging market citizens~\cite{CSCW2021_Razaq} and non-native English speakers~\cite{SOUPS2021_Hasegawa}); and (iii) regarding ``privacy abuse,'' researchers often focus on privacy issues of minorities and demographic groups facing difficulties, one of which is non-Westerners (e.g., women in non-Western countries facing gender gaps~\cite{CHI2019_Sambasivan}).

\section{Discussion}

Our literature-review results demonstrated that UPS has fewer non-Western participant samples than HCI and that the percentage of non-WEIRD participant samples has been slightly decreasing. 
This implies that more significant efforts should be made in UPS to address this geographic diversity issue.
On the basis of our findings, we provide recommendations for further exploring UPS research with geographically diversified populations.

\subsection{Summary of Main Findings}

The main findings of our review and analysis as follows:

\begin{itemize}
\setlength{\itemsep}{5pt}
\setlength{\parskip}{0pt}
\item Only 51.89\% of UPS papers explicitly reported the participants' countries. In non-HCI and non-UPS-focused conferences, authors were less likely to report them. In papers from which we were able to identify or infer them, 78.89\% of the participant samples were Western. The Western skew increased between 2017 and 2021. (RQ1, Sections~\ref{section:results_western} and~\ref{section:results_differences_by_venues})
\item The participant samples tended to come from countries that were considered educated, industrialized, rich, and democratic on the basis of national statistics. The percentages of papers explicitly reporting their participants' education and/or income levels were not high. (RQ1, Sections~\ref{section:results_educated} and~\ref{section:results_industrialized_and_rich})
\item Most of the UPS papers (86.57\%) exclusively comprised authors affiliated with institutions in Western countries. The authors tended to recruit participants from the countries of their affiliations.
When authors recruited participants from other countries, they were more likely to conduct surveys than lab studies or interviews due to linguistic and geographic barriers. (RQ2, Section~\ref{section:results_author})
\item Authors with a non-Western background, social media, and some recruitment methods for developers (e.g., source code repositories and app stores) contribute to the recruitment of non-Westerners. On the other hand, studies on experts excluding developers and studies for demonstrating attack feasibility were skewed heavily toward Western participant samples. (RQ2, Sections~\ref{section:results_author} and~\ref{section:Differences by Research Topics})
\end{itemize}

\subsection{Replications for Non-WEIRD Populations}
\label{section:Replications for Non-WEIRD Populations}

A replication study can be used to examine the generalizability of the findings of an existing study to a new population or to evaluate the robustness of the findings of an existing study when a new study design is used.
Although several existing UPS papers have already shown that regional and cultural differences affect the results~\cite{PETS2017_Li, CHI2023_Herbert, CHI2017_Sawaya}, we found that most UPS papers have examined only WEIRD countries and not considered/discussed whether geographic diversity affects the results. 
Therefore, researchers should explore different insights from geographically diversified populations while pursuing generalizable results.
Recognizing population differences is a promising path for addressing the issues faced by marginalized populations.

Replication studies are recognized to be important and are recommended by conferences dedicated to UPS, for example, SOUPS emphasizes their importance in the Call for Paper (CFP)~\cite{soups_replication}, and USEC and EuroUSEC include them in their CFP topics.
Seven replication papers are in our dataset~\cite{replication_example1,replication_example2,replication_example3,replication_example4,replication_example6,replication_example7,10.1145/3481357.3481511}\footnote{In our dataset, we identified papers with ``Replication:'' in the title as replication papers. SOUPS recommends this title naming rule.}.
However, the current replication studies did not sufficiently focus on non-WEIRD populations; some papers conducted a cross-cultural replication, but only in Western countries.
Replication with an awareness of geographic diversity including non-WEIRD populations should be actively pursued and clearly recommended in the CFP.

The program committee (PC) needs to understand regional, cultural, and institutional differences arising from the geographic diversity to properly evaluate the value of research on non-WEIRD participants and replication studies for non-WEIRD participants. 
For this purpose, Western PC members alone may be insufficient to cover a broad knowledge base. 
In other words, more diversity of PC members is required.
We additionally analyzed the PC diversity in UPS and found that the PC members also largely skewed towards researchers affiliated with institutions in Western
countries: PC members were mostly Western-affiliated researchers (89.44\%, 127/142) at UPS-focused conferences (SOUPS, USEC, and EuroUSEC) and research tracks (``Privacy and Security'' subcommittee in CHI, ``Security Usability and Measurement'' track in CCS) in 2021.
Soneji et al. shed light on the challenges of the review process by conducting interviews with PC members at the top-tier security conferences~\cite{soneji22:peer-review}.
Although they mentioned the issue of PC members' expertise, they did not discuss the impact of the lack of geographic diversity of PC members. 
This implies that the Western skew of PC members is not currently well recognized as a major problem in top-tier security conferences.
UPS-related conferences should be open for PC nominations and select PC members in accordance with geographic diversity. In addition, a complementary approach to enrich PC diversity is actively reaching out to research groups that study non-Western (non-WEIRD) populations and have expertise in cross-cultural/regional research. As a first step, such research groups need not necessarily have a background in non-WEIRD countries as long as they have expertise in cross-cultural/regional research and sufficient cultural knowledge.

\subsection{Reproducibility and Participant Protection}
\label{subsection:Reproducibility in UPS Research}

Papers with low reproducibility may hinder replication studies.
Through our review, we identified many papers with reproducibility problems, i.e., lacking basic demographic information of participants of a user study.
For example, the percentage of papers for which we were able to identify the participants' countries was not sufficiently high (Section~\ref{section:results_western}), nor was the percentage of papers that described the participant's education or income level (Sections~\ref{section:results_educated} and~\ref{section:results_industrialized_and_rich}).
Moreover, most attack-feasibility studies did not report participant demographics, which is vital because many attacks used English audio or input, highlighting the impact of language proficiency (Section~\ref{section:Differences by Research Topics}).
Our recommendation that researchers appropriately report participant demographics can be helpful for researchers who conduct replication studies to analyze generalizability and population differences. 
Note that all participant demographics reported do not necessarily have to be linked to the study results one by one because such brute-force linking entails conducting numerous statistical tests to achieve the desired $p$-value and creates misleading interpretations (called $p$-hacking).

How and where participant demographic information was described were fragmented in many papers, so researchers (readers of these papers) may spend a lot of time searching for the relevant descriptions. 
Moreover, many papers provided some information but not clear and insufficient descriptions (please see frequent examples in Section~\ref{section:results_western}). 
A straightforward approach for researchers to recognize the reporting of participant demographics is using a checklist or uniform format. 
For example, NeurIPS, one of the top AI conferences, has introduced a checklist for reproducibility of AI research~\cite{neurips_checklist}. In addition, Distler et al. suggested a self-check format for UPS researchers to provide a clear understanding of risk representation of user studies~\cite{TOCHI2021_Distler}. 
However, for recognizing and reporting participant demographics, such an approach needs a more careful discussion in the research community because it may (i) simplify sensitive issues in recognizing participant demographics or (ii) create a threat to the safety of at-risk populations.
In the first case, researchers require careful consideration when recognizing certain demographics, e.g., asking transgender people about their recognized gender or asking people living in areas that cannot be clearly defined because of conflict zones or territorial disputes about their countries of residence. 
To avoid these issues, researchers should provide sufficient options for a broad range of individual expressions of identity. 
In the second case, researchers require measures to protect the physical/digital safety of at-risk populations (e.g., political activists and survivors of intimate partner abuse/violence). 
Bellini et al. summarized the practices of protecting at-risk populations' privacy and physical/digital safety against reidentification before, during, or after the research; representative practices include omitting details of research procedures and demographic information~\cite{SoK:SaferDigital-Safety, SEC23_McClearn}. 
Therefore, the UPS community also needs to understand the reasons for not clarifying information associated with participants' identities when participants belong to at-risk populations.

\subsection{Geographic Barriers}
\label{subsection:Issues of Study and Recruitment Methods}

The skew toward Western participant samples in the UPS papers is related to the facts that the majority of authors are affiliated with institutions in Western countries and that authors tend to recruit participants from the countries of their affiliations as mentioned in Section~\ref{section:results_author}.
Geographic barriers may cause researchers to find the most \textit{convenient sample} (e.g., students of the local university who are easy for university researchers to recruit).
Online methods can be used as a promising approach to overcome geographic barriers.
Our results demonstrated that crowdsourcing is the common method to recruit participants outside of authors' countries. 
However, note that there is also a skew in the participant pool in popular crowdsourcing platforms.
Although crowdsourcing platforms such as MTurk, Prolific, and Clickworker are often used to recruit participants as end users, these platforms basically had Western workers, and researchers cannot reach people living in outside of the West, e.g., Asian countries as well as the Global South.
Thus, to overcome the aforementioned geographic barriers of participant recruitment, researchers should consider using local (regional) crowdsourcing platforms used in the country of the target population. 
On the other hand, online methods have a major weakness: only people with Internet access and some familiarity with information technology can participate. 
This is a more serious problem in the Global South. As a complementary solution, we recommend improving the researcher diversity and collaboration with local researchers, as mentioned in Section~\ref{subsection:Researcher diversity}.

Recruiting experts through crowdsourcing is generally more difficult than recruiting non-experts (end users).
This is because most workers registered for crowdsourcing are non-experts.
Even in this situation, software developers are often recruited online, e.g., Freelancer.com (a crowdsourcing platform for freelance developers), contact lists on source code repositories (e.g., GitHub), and app stores (e.g., Google's Play Store) as mentioned in Section~\ref{section:Differences by Research Topics}. 
Unfortunately, however, the current methods of recruiting software developers have been reported to have several contractual and methodological issues.
Danilova et al., whose study was published in 2021, admitted that Freelancer.com rejected their academic tasks of a user study for software developers~\cite{274453}. 
This means a mismatch between researchers' ``academic'' objective and the platform's ``business'' objective. 
Furthermore, while some researchers started recruiting software developers by harvesting developers' emails from source code repositories, app stores, and social media, Tahaei et al. pointed out that there is no support for doing large-scale recruitment by harvesting those platforms (at least on GitHub, this harvesting is against the terms of service and privacy policy)~\cite{10.1145/3491102.3501957}.
In addition, Danilova et al. and Tahaei et al. pointed out the problem that online recruitment often includes software developers who do not have sufficient development skills to participate in the user study, and they evaluated screening methods to verify whether participants have sufficient development skills~\cite{10.1109/ICSE43902.2021.00057,10.1145/3510003.3510223,10.1145/3491102.3501957}.

The majority of studies that recruited experts through online methods (e.g., crowdsourcing, source code repositories, app stores, and social media) focused on software developers. 
However, experts studied in UPS are more diverse and are not limited to software developers, but include system/network administrators, SOC analysts, security experts, and so on.
Since such expert populations are hard to find in the current participant pool of crowdsourcing platforms, recruitment methods tend to rely on personal contacts.
Although there is currently no groundbreaking method for recruiting experts other than software developers, a platform is desired that can be used without geographical or linguistic barriers and designed with incentives to attract professionals.

\subsection{Linguistic Barriers}
\label{subsection:Researcher diversity}

Whereas the online research mentioned in the previous section can (partially) resolve geographic barriers to recruiting, there are also linguistic barriers to studying non-WERID populations.
We suggest approaches to reduce the linguistic barriers in UPS research include (1) enriching the geographic diversity of authors and (2) empowering non-Western researchers.

Although English is used as a national language in many countries, non-native English speakers account for the majority of the world's population, and there are many countries where English proficiency is not sufficiently high~\cite{EFEPI}.
Linguistic barriers resulting from the use of different languages in various countries significantly impact the localization of research within a single linguistic community (Section~\ref{section:results_author}).
Linguistic barriers become apparent following two cases: when (1) participants and researchers and (2) documents and researchers have different native languages.
For the first case of participants and researchers, especially in user studies that require immediate communication, the researchers have difficulty correctly communicating the intention of the study to the participants and to understanding the intended meanings of their responses.
In fact, the results in Section~\ref{section:results_author} suggested that authors were more likely to conduct surveys than lab studies or interviews when they recruited participants from other countries. 
Dell et al. evaluated the difference in participants' response bias due to demand characteristics between a local interviewer and a foreign researcher with a translator~\cite{10.1145/2207676.2208589}. 
Their results demonstrated that the bias towards the artifact of the foreign researcher was greater than that of the local interviewer.
Therefore, the difficulty in conducting interviews with people speaking other languages can be a serious problem in human factor research including UPS.
For the second case of documents and researchers, 
although understanding online public advice, policies (e.g., privacy policies), and user posts on social media\footnote{The papers that analyzed only online public data without a user study were out of the scope of our literature review.} is important work in the UPS field, such documents written in foreign languages are difficult for researchers to correctly analyze.
Redmiles et al. suggested the key challenge in prioritizing security and privacy advice through their large-scale analysis of the advice on the Web~\cite{SEC20_Redmiles}, while the obtained results were inherently constrained to English-language advice owing to the advice collected from MTurkers, mainly consisting of workers from English-speaking countries, including the U.S. 
In their analysis of online phishing advice, Althobaiti et al. explicitly stated the linguistic limitation; they restricted their investigation to webpages from limited countries to avoid heavy reliance on automated translation~\cite{EuroUSEC2019_Althobaiti}. 
When English-speaking researchers conducts a study in a non-English-speaking country, they should preferably work with a researcher who is a native speaker of the language of that country.
From the viewpoint of studies in non-English speaking countries, especially in the Global South, researchers may incur additional costs to translate the participants' responses or documents into the publication language (i.e., English).
To reduce these linguistic barriers, the UPS research community should facilitate cross-national authorship and consider linguistic support, including cost sharing.

The diversification of conference locations is expected to not only facilitate collaboration among researchers by increasing the number of diverse conference attendees~\cite{10.1145/3292013} but also foster and strengthen local (e.g., non-Western) researchers by stimulating local research communities and technology transfers.
In UPS, however, most conferences have been held in limited locations. 
S\&P, NDSS, and SOUPS have never been held outside the U.S., USENIX Security has been held mostly in the U.S. and once in Canada, and CCS has been held almost exclusively in the U.S. and Europe (it was held in South Korea only in 2021).

\subsection{Research Justice Considerations}
Research justice is a concept and method that articulates and ensures fairness, equity, and ethical considerations in research, particularly in community-based collaborations, including non-WEIRD populations. It addresses power imbalances, promotes inclusivity, and respects marginalized populations' or individuals' rights and interests.  
Haelewaters et al. criticized ``helicopter research,'' where Global North (i.e., richer globalized countries such as Western countries) researchers make roundtrips to the Global South to conduct their research with little to no involvement of local collaborators, highlighting it as a prime example of harming research justice~\cite{Haelewaters_TenSimpleRules}.
Haelewaters et al. proposed rules for avoiding helicopter research for better, collaborative, and non-colonial science between the Global North and the Global South. 
The rules mainly include equal and synergistic collaborations with local researchers/collaborators/communities, being ethical and fair about publications and authorship, recognizing differences in work culture, and using local infrastructure. 
These rules are also related to our suggestions: improving researcher diversity and collaboration with local researchers (Sections~\ref{subsection:Issues of Study and Recruitment Methods} and \ref{subsection:Researcher diversity}), fostering and strengthening local researchers by stimulating local research communities and technology transfers (Section~\ref{subsection:Researcher diversity}), increasing the diversity among PC members with knowledge of non-WEIRD societies/populations (Section~\ref{section:Replications for Non-WEIRD Populations}), and using local crowdsourcing platforms (Section~\ref{subsection:Issues of Study and Recruitment Methods}). When facilitating research on non-WEIRD societies/populations, the UPS community needs to embrace Haelewaters et al.'s rules as the principles of research.

\subsection{Future Research}
\label{section:discussion_researchtopics}

In Section~\ref{subsubsection:Harms of WEIRD skew to UPS Research}, we presented how WEIRD skew would harm UPS research. 
Here, we discuss the promising areas of future research in USP, highlighting the undisclosed facts and unsolved challenges  based on our literature reviews.

\noindent\textbf{WEIRD perspective.}
Our results demonstrated that the participant samples were skewed toward Educated, Industrialized, Rich, and Democratic (EIRD) countries as well as Western.
%
With respect to ``Western,'' we found Western skew in UPS papers on any topic. Among these papers, participant samples that investigated experts (excluding developers) were particularly skewed toward Westerners. 
While the papers that focused on developers investigated non-Western participant samples to some extent as mentioned in Section~\ref{section:Differences by Research Topics}, experts are more diverse, including system/network administrators, SOC analysts, and CSIRT engineers.
In software development, organizational structures varied across countries and generated different organizational constraints~\cite{10.1145/3485832.3485922}. 
This fact implies that organizational and cultural issues are a promising direction to study various types of experts in non-Western countries. 
%
With respect to ``Educated,'' studying only educated countries is not sufficient because Global South countries face several challenges in the development of cybersecurity education.
Cybersecurity education in the Global South is generally elementary~\cite{nonW_cybersec_education}, and individual participants' education level was found to be a significant factor influencing participants' security and privacy behaviors in general~\cite{CHI2020_Zou, SOUPS2015_Wash, IFIP2013_Parsons}.
Researchers should clarify what kind of security and privacy education and advice people in the Global South are receiving and make efforts to narrow the digital divide regarding security and privacy education between the Global South and Global North.
%
With respect to ``Industrialized'' and ``Rich,'' some studies have already focused on the security and privacy concerns of people in the Global South, where people have limited IT resources, e.g., concerns regarding device sharing, and developed support technology for them~\cite{SOUPS2017_Baqer, CHI2019_Ahmed, CHI2018_Robinson}.
Researchers should investigate whether support technology is actionable in other countries and regions with lower levels of industrialization and wealth.
%
With respect to ``Democratic,'' some studies indicated that people in countries with little political freedom perceive their privacy differently than people in other countries~\cite{democratic_privacy}. 
This democratic difference may affect the study result of the topic of ``access control and privacy preferences'': participant samples of this topic were particularly skewed toward Westerners as mentioned in Section~\ref{section:Differences by Research Topics}.
When conducting a user study with potential participants in politically repressive regimes, researchers should consider and overcome difficulties in recruiting methods and study methods, as well as the ethical issue of potentially putting participants at risk due to their criticism of the government.
The research practices for protecting such at-risk participants are introduced in Section~\ref{subsection:Reproducibility in UPS Research}.

\noindent\textbf{Design evaluation.}
Although some cross-cultural comparisons of user perceptions, concerns, and practices have been conducted in the UPS field, those of design evaluation have been insufficiently investigated.
In the HCI and psychology fields, user preferences on interface design and framing are known to vary across countries~\cite{framing_cultural, interface_cultural1, interface_cultural2}; therefore, more differences across countries in the usability of security/privacy interfaces and the effectiveness of framing, such as text and icons on the notification that encourage users to adopt a security technology, should be identified to design interfaces tailored to each country's users.

\noindent \textbf{More diverse non-WEIRD societies/populations.}
Broadening geographic diversity is a vital factor in fostering UPS growth; nevertheless, there is a pitfall in the measure of the percentage of non-Western participant samples. 
If more papers studied India and China, which have already been studied to some extent (the fifth and sixth highest number of participant samples, respectively), the percentage of non-Western participant samples would increase; however, there are still many non-Western countries in Africa, South America, and Asia that have not been studied in the last five years (please see Section~\ref{section:results_western}).
The UPS research community should strive to understand more diverse countries.

\subsection{Limitations}

\noindent
\textbf{Measurement errors.}
While some measurement errors are inevitable in manual analysis, we have tried to make our literature-review procedure as clear as possible to ensure coding precision.
Our dataset will be available to researchers upon request to improve reproducibility and facilitate replications of our study.

\noindent
\textbf{Conference coverage.}
Our dataset (ten conferences, $N$=715) is much larger than the recent representative literature-review papers in UPS (five conferences, $N$=284 in~\cite{TOCHI2021_Distler}). Our dataset covers top-tier and competitive conferences in which UPS papers are presented and that highly influence many researchers and societies globally. Understanding the WEIRD skew of such conferences would be a reasonable first step. Note that papers concerning non-WEIRD countries are also presented at region-focused conferences (e.g., AfriCHI), which are not included in our dataset.

\noindent
\textbf{Exclusion of papers that analyzed online public data.}
We focused on the UPS papers in which the authors recruited participants for user studies, and we did not focus on the UPS papers in which the authors analyzed online public data.
We are concerned that such UPS papers are also skewed towards Western societies, i.e., researchers investigated design and policies implemented on Western online services and online content posted by Western users.

\noindent
\textbf{Limitation of WEIRD framework.} 
Most indicators of the WEIRD framework\footnote{Indicators of E and I use both national averages and individual figures.} assume representative population demographics in each country. 
However, in general, individuals within the same country are diverse. 
For example, even if the participants are from poor countries, they may be personally wealthy. 
Moreover, online methods can easily recruit people with Internet access and some familiarity with information technology, as mentioned in Section~\ref{subsection:Issues of Study and Recruitment Methods}.
In addition, low-income people in Western countries remain valuable for the UPS community. However, we were unable to quantify the papers on such people because few papers reported their participants' income levels.

\noindent
\textbf{Other indicators of the diversity dimensions.}
The five indicators of WEIRD are just a subset of the diversity dimensions that are considered in human factor research~\cite{CHI19_Himmelsbach}.
UPS researchers should consider various kinds of diversities beyond WEIRD.
For example, Hofstede's cultural dimensions framework includes promising indicators. 
We additionally analyzed our dataset in terms of that framework (please see Appendix~\ref{section:appendix_Hofstede}). 
Future research is needed to comprehensively capture the diversities that influence UPS research.

\section{Conclusion}

The goal of this study was to understand the geographic diversity of recruited participants in usable privacy and security (UPS) and the characteristics of the methodologies and research topics. 
Throughout our comprehensive literature review of 715 UPS papers, we unveiled that UPS papers are highly skewed towards Western participant samples (78.89\%). 
Unfortunately, this skew is higher than that in human-computer interaction (HCI; 73.13\%). 
Moreover, the percentage of non-Western participant samples in UPS papers between 2017–2021 decreased (25.00\% to 20.77\%, 0.83x).
This implies that more significant efforts should be made in UPS to address this geographic diversity issue. 
We provided the recommendations for making UPS research less WEIRD (Western, Educated, Industrialized, Rich, and Democratic): facilitating replications, addressing geographic and linguistic issues of study and recruiting methods, and setting future research topics.
We hope that our findings and recommendations encourage the growing UPS community to further diversify and become more inclusive.

\section*{Acknowledgements}
We would like to thank Sebastian Linxen, the author of the original study (the Linxen study)~\cite{CHI2021_Linxen}, for the advice on our replication efforts.
We also thank the anonymous reviewers and shepherd for their insightful and invaluable feedback.

\bibliographystyle{plain}
\bibliography{ref}

\begin{thebibliography}{100}

\bibitem{neurips_checklist}
NeurIPS 2021.
\newblock Paper checklist guidelines.
\newblock \url{https://neurips.cc/Conferences/2021/PaperInformation/PaperChecklist}, (accessed August 2, 2022).

\bibitem{soups_replication}
SOUPS 2022.
\newblock Call for papers.
\newblock \url{https://www.usenix.org/conference/soups2022/call-for-papers}, (accessed August 2, 2022).

\bibitem{10.1145/3411764.3445122}
Noura Abdi, Xiao Zhan, Kopo~M. Ramokapane, and Jose Such.
\newblock Privacy norms for smart home personal assistants.
\newblock In {\em Proceedings of the 2021 CHI Conference on Human Factors in Computing Systems}, 2021.

\bibitem{SP2017_Abu}
Ruba Abu-Salma, M.~Angela Sasse, Joseph Bonneau, Anastasia Danilova, Alena Naiakshina, and Matthew Smith.
\newblock Obstacles to the adoption of secure communication tools.
\newblock In {\em Proceedings of the 38th IEEE Symposium on Security and Privacy}, 2017.

\bibitem{10.1207/S15327051HCI1523_5}
Mark~S. Ackerman.
\newblock The intellectual challenge of {CSCW}: The gap between social requirements and technical feasibility.
\newblock {\em Hum.-Comput. Interact.}, 15(2):179–203, 2000.

\bibitem{CSUR2017_Acquisti}
Alessandro Acquisti, Idris Adjerid, Rebecca Balebako, Laura Brandimarte, Lorrie~Faith Cranor, Saranga Komanduri, Pedro~Giovanni Leon, Norman Sadeh, Florian Schaub, Manya Sleeper, Yang Wang, and Shomir Wilson.
\newblock Nudges for privacy and security: Understanding and assisting users' choices online.
\newblock {\em ACM Computing Surveys}, 50(3):1--41, 2017.

\bibitem{10.1145/322796.322806}
Anne Adams and Martina~Angela Sasse.
\newblock Users are not the enemy.
\newblock {\em Commun. ACM}, 42(12):40–46, 1999.

\bibitem{CSCW2017_Ahmed}
Syed~Ishtiaque Ahmed, Md~Romael Haque, Jay Chen, and Nicola Dell.
\newblock Digital privacy challenges with shared mobile phone use in {B}angladesh.
\newblock {\em Human-Computer Interaction}, 1(CSCW):1--20, 2017.

\bibitem{CHI2017_Ahmed}
Syed~Ishtiaque Ahmed, Md~Romael Haque, Shion Guha, Md~Rashidujjaman Rifat, and Nicola Dell.
\newblock Privacy, security, and surveillance in the {G}lobal {S}outh: A study of biometric mobile {SIM} registration in {B}angladesh.
\newblock In {\em Proceedings of the 2017 CHI Conference on Human Factors in Computing Systems}, 2017.

\bibitem{CHI2019_Ahmed}
Syed~Ishtiaque Ahmed, Md~Romael Haque, Irtaza Haider, Jay Chen, and Nicola Dell.
\newblock ``{E}veryone has some personal stuff'': {D}esigning to support digital privacy with shared mobile phone use in {B}angladesh.
\newblock In {\em Proceedings of the 2019 CHI Conference on Human Factors in Computing Systems}, 2019.

\bibitem{EuroUSEC2020_Qahtani}
Elham Al~Qahtani, Yousra Javed, Heather Lipford, and Mohamed Shehab.
\newblock Do women in conservative societies (not) follow smartphone security advice? {A} case study of {S}audi {A}rabia and {P}akistan.
\newblock In {\em Proceedings of the 5th European Workshop on Usable Security}, 2020.

\bibitem{10.1145/3411764.3445574}
Kholoud Althobaiti, Nicole Meng, and Kami Vaniea.
\newblock I don't need an expert! {M}aking {URL} phishing features human comprehensible.
\newblock In {\em Proceedings of the 2021 CHI Conference on Human Factors in Computing Systems}, 2021.

\bibitem{EuroUSEC2019_Althobaiti}
Kholoud Althobaiti, Ghaidaa Rummani, and Kami Vaniea.
\newblock A review of human-and computer-facing {URL} phishing features.
\newblock In {\em Proceedings of the 4th European Workshop on Usable Security}, 2019.

\bibitem{AP2008_Arnett}
Jeffrey~J. Arnett.
\newblock The neglected 95\%: Why {A}merican psychology needs to become less {A}merican.
\newblock {\em American Psychologist}, 63(7):602–614, 2008.

\bibitem{replication_example7}
Khadija Baig, Elisa Kazan, Kalpana Hundlani, Sana Maqsood, and Sonia Chiasson.
\newblock Replication: Effects of media on the mental models of technical users.
\newblock In {\em Proceedings of the 17th Symposium on Usable Privacy and Security}, 2021.

\bibitem{SOUPS2017_Baqer}
Khaled Baqer, Ross Anderson, Jeunese~Adrienne Payne, Lorna Mutegi, and Joseph Sevilla.
\newblock Digitally: {P}iloting offline payments for phones.
\newblock In {\em Proceedings of the 13th Symposium on Usable Privacy and Security}, 2017.

\bibitem{SoK:SaferDigital-Safety}
Rosanna Bellini, Emily Tseng, Noel Warford, Tara~Matthews Alaa~Daffalla, Sunny Consolvo, Jill~Palzkill Woelfer, Patrick~Gage Kelley, Michelle~L. Mazurek, Dana Cuomo, Nicola Dell, and Thomas Ristenpart.
\newblock So{K}: Safer digital-safety research involving at-risk users.
\newblock {\em arXiv: 2309.00735}, 2023.

\bibitem{replication_example4}
Karoline Busse, Julia Sch{\"a}fer, and Matthew Smith.
\newblock Replication: No one can hack my mind revisiting a study on expert and non-expert security practices and advice.
\newblock In {\em Proceedings of the 15th Symposium on Usable Privacy and Security}, 2019.

\bibitem{replication_example1}
Karoline Busse, Dominik Wermke, Sabrina Amft, Sascha Fahl, Emanuel von Zezschwitz, and Matthew Smith.
\newblock Replication: Do we snooze if we can't lose? {M}odelling risk with incentives in habituation user studies.
\newblock In {\em Proceedings of the 2019 Workshop on Usable Security}, 2019.

\bibitem{HAISA2017_Butavicius}
Marcus~A. Butavicius, Kathryn Parsons, Malcolm~R. Pattinson, Agata McCormac, Dragana Calic, and Meredith Lillie.
\newblock Understanding susceptibility to phishing emails: Assessing the impact of individual differences and culture.
\newblock In {\em Proceedings of the 11th International Symposium on Human Aspects of Information Security \& Assurance}, 2017.

\bibitem{replication_example2}
Casey Canfield, Alex Davis, Baruch Fischhoff, Alain Forget, Sarah Pearman, and Jeremy Thomas.
\newblock Replication: Challenges in using data logs to validate phishing detection ability metrics.
\newblock In {\em Proceedings of the 13th Symposium on Usable Privacy and Security}, 2017.

\bibitem{SEC2021_Cao2021}
Weicheng Cao, Chunqiu Xia, Sai~Teja Peddinti, David Lie, Nina Taft, and Lisa~M Austin.
\newblock A large scale study of user behavior, expectations and engagement with {A}ndroid permissions.
\newblock In {\em Proceedings of the 30th USENIX Security Symposium}, 2021.

\bibitem{nonW_cybersec_education}
Frankie~E. Catota, M.~Granger Morgan, and Douglas~C. Sicker.
\newblock Cybersecurity education in a developing nation: The ecuadorian environment.
\newblock {\em Journal of Cybersecurity}, 5(1), 2019.

\bibitem{criticism1}
Giacomo Chiozza.
\newblock Is there a clash of civilizations? {E}vidence from patterns of international conflict involvement, 1946-97.
\newblock {\em Journal of Peace Research}, 39(6):711--734, 2002.

\bibitem{EM2016_Cockcroft}
Sophie Cockcroft and Saphira Rekker.
\newblock The relationship between culture and information privacy policy.
\newblock {\em Electronic Markets}, 26(1):55--72, 2016.

\bibitem{law_bycountry1}
comforte AG.
\newblock 17 countries with {GDPR}-like data privacy laws.
\newblock \url{https://insights.comforte.com/countries-with-gdpr-like-data-privacy-laws}, (accessed June 2, 2023).

\bibitem{10.1145/3510003.3510223}
Anastasia Danilova, Stefan Horstmann, Matthew Smith, and Alena Naiakshina.
\newblock Testing time limits in screener questions for online surveys with programmers.
\newblock In {\em Proceedings of the 44th International Conference on Software Engineering}, 2022.

\bibitem{replication_example6}
Anastasia Danilova, Alena Naiakshina, Johanna Deuter, and Matthew Smith.
\newblock Replication: On the ecological validity of online security developer studies: Exploring deception in a password-storage study with freelancers.
\newblock In {\em Proceedings of the 16th Symposium on Usable Privacy and Security}, 2020.

\bibitem{10.1109/ICSE43902.2021.00057}
Anastasia Danilova, Alena Naiakshina, Stefan Horstmann, and Matthew Smith.
\newblock Do you really code? {D}esigning and evaluating screening questions for online surveys with programmers.
\newblock In {\em Proceedings of the 43rd International Conference on Software Engineering}, 2021.

\bibitem{274453}
Anastasia Danilova, Alena Naiakshina, Anna Rasgauski, and Matthew Smith.
\newblock Code reviewing as methodology for online security studies with developers - {A} case study with freelancers on password storage.
\newblock In {\em Proceedings of the 17th Symposium on Usable Privacy and Security}, 2021.

\bibitem{10.1145/2207676.2208589}
Nicola Dell, Vidya Vaidyanathan, Indrani Medhi, Edward Cutrell, and William Thies.
\newblock ``{Y}ours is better!'': {P}articipant response bias in {HCI}.
\newblock In {\em Proceedings of the 2012 SIGCHI Conference on Human Factors in Computing Systems}, 2012.

\bibitem{TOCHI2021_Distler}
Verena Distler, Matthias Fassl, Hana Habib, Katharina Krombholz, Gabriele Lenzini, Carine Lallemand, Lorrie~Faith Cranor, and Vincent Koenig.
\newblock A systematic literature review of empirical methods and risk representation in usable privacy and security research.
\newblock {\em ACM Transactions on Computer-Human Interaction}, 28(6):1--50, 2021.

\bibitem{EFEPI}
EF~Education First.
\newblock E{F} english proficiency index.
\newblock \url{https://www.ef.com/wwen/epi/}, (accessed August 12, 2022).

\bibitem{IRR}
Joseph~L. Fleiss, Bruce Levin, and Myunghee~Cho Paik.
\newblock The measurement of interrater agreement.
\newblock {\em Statistical methods for rates and proportions}, 2(212-236):22--23, 1981.

\bibitem{SOUPS2021_Franz}
Anjuli Franz, Verena Zimmermann, Gregor Albrecht, Katrin Hartwig, Christian Reuter, Alexander Benlian, and Joachim Vogt.
\newblock {S}o{K}: Still plenty of phish in the sea -- {A} taxonomy of user-oriented phishing interventions and avenues for future research.
\newblock In {\em Proceedings of the 17th Symposium on Usable Privacy and Security}, 2021.

\bibitem{criticism2}
George Fujii, Andrew Szarejko, and Diane Labrosse.
\newblock The clash of civilizations in the {IR} classroom.
\newblock \url{https://networks.h-net.org/node/28443/discussions/5273269/h-diploissf-teaching-roundtable-11-6-clash-civilizations-ir}, (accessed November 27, 2022).

\bibitem{SLISPT2014_Garfinkel}
Simson Garfinkel and Heather~Richter Lipford.
\newblock Usable security: History, themes, and challenges.
\newblock {\em Synthesis Lectures on Information Security, Privacy, and Trust}, 5(2):1--124, 2014.

\bibitem{GDP1}
The World~Bank Group.
\newblock {GDP} per capita, {PPP} (current international \$) (2020).
\newblock \url{https://data.worldbank.org/indicator/NY.GDP.PCAP.PP.CD}.

\bibitem{GNI1}
The World~Bank Group.
\newblock {GNI} per capita, {PPP} (current international \$) (2020).
\newblock \url{https://data.worldbank.org/indicator/NY.GNP.PCAP.PP.CD}.

\bibitem{CSranking}
Guofei Gu.
\newblock Computer security conference ranking and statistic.
\newblock \url{https://people.engr.tamu.edu/guofei/sec_conf_stat.htm}, (accessed August 18, 2022).

\bibitem{Haelewaters_TenSimpleRules}
Danny Haelewaters, Tina~A. Hofmann, and Adriana~L. Romero-Olivares.
\newblock Ten simple rules for {G}lobal {N}orth researchers to stop perpetuating helicopter research in the {G}lobal {S}outh.
\newblock {\em PLoS Computational Biology}, 17(8), 2021.

\bibitem{SOUPS2021_Hasegawa}
Ayako~A. Hasegawa, Naomi Yamashita, Mitsuaki Akiyama, and Tatsuya Mori.
\newblock Why they ignore {E}nglish emails: The challenges of non-{N}ative speakers in identifying phishing emails.
\newblock In {\em Proceedings of the 17th Symposium on Usable Privacy and Security}, 2021.

\bibitem{BBS2010_Henrich}
Joseph Henrich, Steven~J. Heine, and Ara Norenzayan.
\newblock The weirdest people in the world?
\newblock {\em Behavioral and Brain Sciences}, 33(2-3):61--83, 2010.

\bibitem{CHI2023_Herbert}
Franziska Herbert, Steffen Becker, Leonie Schaewitz, Jonas Hielscher, Marvin Kowalewski, Angela Sasse, Yasemin Acar, and Markus D{\"u}rmuth.
\newblock A world full of privacy and security (mis)conceptions? {F}indings of a representative survey in 12 countries.
\newblock In {\em Proceedings of the 2023 CHI Conference on Human Factors in Computing Systems}, 2023.

\bibitem{CHI19_Himmelsbach}
Julia Himmelsbach, Stephanie Schwarz, Cornelia Gerdenitsch, Beatrix Wais-Zechmann, Jan Bobeth, and Manfred Tscheligi.
\newblock Do we care about diversity in human computer interaction: {A} comprehensive content analysis on diversity dimensions in research.
\newblock In {\em Proceedings of the 2019 CHI Conference on Human Factors in Computing Systems}, 2019.

\bibitem{hofstede}
Geert Hofstede.
\newblock Dimensionalizing cultures: The {H}ofstede model in context.
\newblock {\em Online readings in psychology and culture}, 2(1):2307--0919, 2011.

\bibitem{politicalrights}
Freedom House.
\newblock Freedom in the world (2020).
\newblock \url{https://freedomhouse.org/report/freedom-world}.

\bibitem{Huntington}
Samuel~P. Huntington.
\newblock {\em The clash of civilizations and the remaking of world order}.
\newblock Simon \& Schuster, 2011.

\bibitem{law_bycountry2}
Enzuzo Inc.
\newblock Data privacy laws in 2023: The rules \& regulations you need to know.
\newblock \url{https://www.enzuzo.com/blog/data-privacy-laws}, (accessed June 2, 2023).

\bibitem{hofstede_website}
Hofstede Insights.
\newblock Country comparison tool.
\newblock \url{https://www.hofstede-insights.com/}, (accessed August 2, 2022).

\bibitem{interface_cultural2}
Jainaba Jagne and Serengul Smith.
\newblock Cross-cultural interface design strategy.
\newblock {\em Universal Access in the Information Society}, 5:299--305, 2006.

\bibitem{10.1145/3485832.3485922}
Fumihiro Kanei, Ayako~Akiyama Hasegawa, Eitaro Shioji, and Mitsuaki Akiyama.
\newblock A cross-role and bi-national analysis on security efforts and constraints of software development projects.
\newblock In {\em Proceedings of the 37th Annual Computer Security Applications Conference}, 2021.

\bibitem{arXiv2021_Kaur}
Mannat Kaur, Michel van Eeten, Marijn Janssen, Kevin Borgolte, and Tobias Fiebig.
\newblock Human factors in security research: Lessons learned from 2008-2018.
\newblock {\em arXiv:2103.13287}, 2021.

\bibitem{JGITM2019_Kshetri}
Nir Kshetri.
\newblock Cybercrime and cybersecurity in {A}frica.
\newblock {\em Journal of Global Information Technology Management}, 22(2):77--81, 2019.

\bibitem{SOUPS2013_Leon}
Pedro~Giovanni Leon, Blase Ur, Yang Wang, Manya Sleeper, Rebecca Balebako, Richard Shay, Lujo Bauer, Mihai Christodorescu, and Lorrie~Faith Cranor.
\newblock What matters to users? {F}actors that affect users' willingness to share information with online advertisers.
\newblock In {\em Proceedings of the 9th symposium on usable privacy and security}, 2013.

\bibitem{GDP2}
Philipp Lepenies.
\newblock {\em The power of a single number: {A} political history of {GDP}}.
\newblock Columbia University Press, 2016.

\bibitem{IJIM2019_Li}
Ling Li, Wu~He, Li~Xu, Ivan Ash, Mohd Anwar, and Xiaohong Yuan.
\newblock Investigating the impact of cybersecurity policy awareness on employees’ cybersecurity behavior.
\newblock {\em International Journal of Information Management}, 45:13--24, 2019.

\bibitem{CSCW2021_Li}
Yao Li, Reza Ghaiumy~Anaraky, and Bart Knijnenburg.
\newblock How not to measure social network privacy: A cross-country investigation.
\newblock {\em Human-Computer Interaction}, 5(CSCW):1--32, 2021.

\bibitem{PETS2017_Li}
Yao Li, Alfred Kobsa, Bart~P. Knijnenburg, and M-H~Carolyn Nguyen.
\newblock Cross-cultural privacy prediction.
\newblock In {\em Proceedings of the 17th Privacy Enhancing Technologies Symposium}, PETS'17, 2017.

\bibitem{10.1145/3134702}
Yifang Li, Nishant Vishwamitra, Bart~P. Knijnenburg, Hongxin Hu, and Kelly Caine.
\newblock Effectiveness and users' experience of obfuscation as a privacy-enhancing technology for sharing photos.
\newblock {\em Human-Computer Interaction}, 1(CSCW):1--24, 2017.

\bibitem{CHI2021_Linxen}
Sebastian Linxen, Christian Sturm, Florian Br{\"u}hlmann, Vincent Cassau, Klaus Opwis, and Katharina Reinecke.
\newblock How {WEIRD} is {CHI}?
\newblock In {\em Proceedings of the 2021 CHI conference on human factors in computing systems}, 2021.

\bibitem{interface_cultural1}
Aaron Marcus and Emilie~West Gould.
\newblock Crosscurrents: Cultural dimensions and global web user-interface design.
\newblock {\em Interactions}, 7(4):32–46, 2000.

\bibitem{10.1145/3290605.3300819}
Diogo Marques, Tiago Guerreiro, Luis Carri\c{c}o, Ivan Beschastnikh, and Konstantin Beznosov.
\newblock Vulnerability \& blame: Making sense of unauthorized access to smartphones.
\newblock In {\em Proceedings of the 2019 CHI Conference on Human Factors in Computing Systems}, 2019.

\bibitem{WIPS2022_Mathew}
Raima Mathew, Mutmaina Adebayo, and Camille Cobb.
\newblock A meta-analysis on the diversity of security and privacy research.
\newblock In {\em Proceedings of the 7th Workshop on Inclusive Privacy and Security}, 2022.

\bibitem{10.1145/3025453.3025875}
Tara Matthews, Kathleen O'Leary, Anna Turner, Manya Sleeper, Jill~Palzkill Woelfer, Martin Shelton, Cori Manthorne, Elizabeth~F. Churchill, and Sunny Consolvo.
\newblock Stories from survivors: Privacy \& security practices when coping with intimate partner abuse.
\newblock In {\em Proceedings of the 2017 CHI Conference on Human Factors in Computing Systems}, 2017.

\bibitem{SEC23_McClearn}
Jessica McClearn, Rikke~Bjerg Jensen, and Reem Talhouk.
\newblock Othered, silenced and scapegoated: Understanding the situated security of marginalised populations in lebanon.
\newblock In {\em Proceedings of the 32nd USENIX Security Symposium}, 2023.

\bibitem{10.1145/3292013}
Helena Mentis, Cliff Lampe, Regina Bernhaupt, Anirudha Joshi, Susan Fussell, Susan Dray, Dan Olsen, Aaron Quigley, Julie~R. Williamson, Eunice Sari, Loren Terveen, Allison Druin, and Philippe Palanque.
\newblock The new {SIGCHI} {EC}'s values and strategic initiatives.
\newblock {\em Interactions}, 26(1):84–85, 2018.

\bibitem{CACM1995_Milberg}
Sandra~J Milberg, Sandra~J Burke, H~Jeff Smith, and Ernest~A Kallman.
\newblock Values, personal information privacy, and regulatory approaches.
\newblock {\em Communications of the ACM}, 38(12):65--74, 1995.

\bibitem{10.1145/3313831.3376791}
Alena Naiakshina, Anastasia Danilova, Eva Gerlitz, and Matthew Smith.
\newblock On conducting security developer studies with {CS} students: Examining a password-storage study with {CS} students, freelancers, and company developers.
\newblock In {\em Proceedings of the 2020 CHI Conference on Human Factors in Computing Systems}, 2020.

\bibitem{framing_cultural}
Yeseul Nam, Haeyoung~Gideon Park, and Young-Hoon Kim.
\newblock Do you favor positive information or dislike negative information? {C}ultural variations in the derivation of the framing effect.
\newblock {\em Current Psychology}, 2021.

\bibitem{worldpopulation}
United Nations.
\newblock World population prospects 2019 (2020 estimation).
\newblock \url{https://population.un.org/wpp/DataQuery/}.

\bibitem{NIST_SP800}
NIST.
\newblock {NIST} {SP}-800 series.
\newblock \url{https://csrc.nist.gov/publications/sp800}, (accessed June 2, 2023).

\bibitem{CHI2020_Nouwens}
Midas Nouwens, Ilaria Liccardi, Michael Veale, David Karger, and Lalana Kagal.
\newblock Dark patterns after the {GDPR}: Scraping consent pop-ups and demonstrating their influence.
\newblock In {\em Proceedings of the 2020 CHI conference on human factors in computing systems}, 2020.

\bibitem{SP2022_Obada}
Borke Obada-Obieh, Yue Huang, Lucrezia Spagnolo, and Konstantin Beznosov.
\newblock {S}o{K}: The dual nature of technology in sexual abuse.
\newblock In {\em Proceedings of the 43rd IEEE Symposium on Security and Privacy}, 2022.

\bibitem{10.1145/3025453.3025831}
Daniela Oliveira, Harold Rocha, Huizi Yang, Donovan Ellis, Sandeep Dommaraju, Melis Muradoglu, Devon Weir, Adam Soliman, Tian Lin, and Natalie Ebner.
\newblock Dissecting spear phishing emails for older vs young adults: On the interplay of weapons of influence and life domains in predicting susceptibility to phishing.
\newblock In {\em Proceedings of the 2017 CHI Conference on Human Factors in Computing Systems}, 2017.

\bibitem{workercountries2}
Jonas Oppenlaender, Kristy Milland, Aku Visuri, Panos Ipeirotis, and Simo Hosio.
\newblock Creativity on paid crowdsourcing platforms.
\newblock In {\em Proceedings of the 2020 CHI Conference on Human Factors in Computing Systems}, 2020.

\bibitem{IFIP2013_Parsons}
Kathryn Parsons, Agata McCormac, Malcolm Pattinson, Marcus Butavicius, and Cate Jerram.
\newblock Phishing for the truth: A scenario-based experiment of users' behavioural response to emails.
\newblock In {\em Proceedings of the 2013 IFIP international information security conference}, 2013.

\bibitem{238303}
Sarah Pearman, Shikun~Aerin Zhang, Lujo Bauer, Nicolas Christin, and Lorrie~Faith Cranor.
\newblock Why people (don't) use password managers effectively.
\newblock In {\em Proceedings of the 15th Symposium on Usable Privacy and Security}, 2019.

\bibitem{workercountries1}
Eyal Peer, Laura Brandimarte, Sonam Samat, and Alessandro Acquisti.
\newblock Beyond the {T}urk: Alternative platforms for crowdsourcing behavioral research.
\newblock {\em Journal of Experimental Social Psychology}, 70:153--163, 2017.

\bibitem{education_stats}
United Nations~Development Programme.
\newblock Human development report 2019.
\newblock \url{https://hdr.undp.org/content/human-development-report-2019}, (accessed August 2, 2022).

\bibitem{CSCW2021_Razaq}
Lubna Razaq, Tallal Ahmad, Samia Ibtasam, Umer Ramzan, and Shrirang Mare.
\newblock ``{W}e even borrowed money from our neighbor'': {U}nderstanding mobile-based frauds through victims' experiences.
\newblock {\em Human-Computer Interaction}, 5(CSCW):1--30, 2021.

\bibitem{SP2019_Redmiles}
Elissa~M. Redmiles, Sean Kross, and Michelle~L Mazurek.
\newblock How well do my results generalize? {C}omparing security and privacy survey results from {MT}urk, web, and telephone samples.
\newblock In {\em Proceedings of the 40th IEEE Symposium on Security and Privacy}, 2019.

\bibitem{SEC20_Redmiles}
Elissa~M. Redmiles, Noel Warford, Amritha Jayanti, Aravind Koneru, Sean Kross, Miraida Morales, Rock Stevens, and Michelle~L Mazurek.
\newblock A comprehensive quality evaluation of security and privacy advice on the web.
\newblock In {\em Proceedings of the 29th USENIX Security Symposium}, 2020.

\bibitem{SOUPS2020_Reinheimer}
Benjamin Reinheimer, Lukas Aldag, Peter Mayer, Mattia Mossano, Reyhan Duezguen, Bettina Lofthouse, Tatiana Von~Landesberger, and Melanie Volkamer.
\newblock An investigation of phishing awareness and education over time: When and how to best remind users.
\newblock In {\em Proceedings of the 16th Symposium on Usable Privacy and Security}, 2020.

\bibitem{CHI2018_Robinson}
Simon Robinson, Jennifer Pearson, Thomas Reitmaier, Shashank Ahire, and Matt Jones.
\newblock Make yourself at phone: Reimagining mobile interaction architectures with emergent users.
\newblock In {\em Proceedings of the 2018 CHI Conference on Human Factors in Computing Systems}, 2018.

\bibitem{CHI2019_Sambasivan}
Nithya Sambasivan, Amna Batool, Nova Ahmed, Tara Matthews, Kurt Thomas, Laura~Sanely Gayt{\'a}n-Lugo, David Nemer, Elie Bursztein, Elizabeth Churchill, and Sunny Consolvo.
\newblock ``{T}hey don't leave us alone anywhere we go'': {G}ender and digital abuse in {S}outh {A}sia.
\newblock In {\em Proceedings of the 2019 CHI Conference on Human Factors in Computing Systems}, 2019.

\bibitem{SOUPS2018_Sambasivan}
Nithya Sambasivan, Garen Checkley, Amna Batool, Nova Ahmed, David Nemer, Laura~Sanely Gayt{\'a}n-Lugo, Tara Matthews, Sunny Consolvo, and Elizabeth~F. Churchill.
\newblock ``{P}rivacy is not for me, it's for those rich women'': {P}erformative privacy practices on mobile phones by women in {S}outh {A}sia.
\newblock In {\em Proceedings of the 14th Symposium on Usable Privacy and Security}, 2018.

\bibitem{CHI2017_Sawaya}
Yukiko Sawaya, Mahmood Sharif, Nicolas Christin, Ayumu Kubota, Akihiro Nakarai, and Akira Yamada.
\newblock Self-confidence trumps knowledge: {A} cross-cultural study of security behavior.
\newblock In {\em Proceedings of the 2017 CHI Conference on Human Factors in Computing Systems}, 2017.

\bibitem{10.1145/3481357.3481511}
Maxim Schessler, Eva Gerlitz, Maximilian H\"{a}ring, and Matthew Smith.
\newblock Replication: Measuring user perceptions in smartphone security and privacy in {G}ermany.
\newblock In {\em Proceedings of the 6th European Workshop on Usable Security}, 2021.

\bibitem{263788}
Bingyu Shen, Lili Wei, Chengcheng Xiang, Yudong Wu, Mingyao Shen, Yuanyuan Zhou, and Xinxin Jin.
\newblock Can systems explain permissions better? {U}nderstanding users' misperceptions under smartphone runtime permission model.
\newblock In {\em Proceedings of the 30th USENIX Security Symposium}, 2021.

\bibitem{pdfminer}
Yusuke Shinyama, Philippe Guglielmetti, and Pieter Marsman.
\newblock pdfminer.six.
\newblock \url{https://pdfminersix.readthedocs.io/en/latest/}, (accessed June 12, 2022).

\bibitem{255682}
Justin Smith, Lisa Nguyen~Quang Do, and Emerson Murphy-Hill.
\newblock Why can't {J}ohnny fix vulnerabilities: A usability evaluation of static analysis tools for security.
\newblock In {\em Proceedings of the 16th Symposium on Usable Privacy and Security}, 2020.

\bibitem{soneji22:peer-review}
Ananta Soneji, Faris~Bugra Kokulu, Carlos Rubio-Medrano, Tiffany Bao, Ruoyu Wang, Yan Shoshitaishvili, and Adam Doup\'e.
\newblock ``{F}lawed, but like democracy we don't have a better system'': The experts' insights on the peer review process of evaluating security papers.
\newblock In {\em Proceedings of the 43rd IEEE Symposium on Security and Privacy}, 2022.

\bibitem{PETS2021_Story}
Peter Story, Daniel Smullen, Yaxing Yao, Alessandro Acquisti, Lorrie~Faith Cranor, Norman Sadeh, and Florian Schaub.
\newblock Awareness, adoption, and misconceptions of web privacy tools.
\newblock 2021.

\bibitem{JGIM2002_Straub}
Detmar Straub, Karen Loch, Roberto Evaristo, Elena Karahanna, and Mark Srite.
\newblock Toward a theory-based measurement of culture.
\newblock {\em Journal of Global Information Management}, 10(1):13--23, 2002.

\bibitem{JCMC2014_Taddicken}
Monika Taddicken.
\newblock The `privacy paradox' in the social web: The impact of privacy concerns, individual characteristics, and the perceived social relevance on different forms of self-disclosure.
\newblock {\em Journal of computer-mediated communication}, 19(2):248--273, 2014.

\bibitem{EuroUSEC2019_Tahaei}
Mohammad Tahaei and Kami Vaniea.
\newblock A survey on developer-centred security.
\newblock In {\em Proceedings of the 4th European Workshop on Usable Security}, 2019.

\bibitem{10.1145/3491102.3501957}
Mohammad Tahaei and Kami Vaniea.
\newblock Recruiting participants with programming skills: {A} comparison of four crowdsourcing platforms and a {CS} student mailing list.
\newblock In {\em Proceedings of the 2022 CHI Conference on Human Factors in Computing Systems}, 2022.

\bibitem{10.1145/3372297.3417882}
Joshua Tan, Lujo Bauer, Nicolas Christin, and Lorrie~Faith Cranor.
\newblock Practical recommendations for stronger, more usable passwords combining minimum-strength, minimum-length, and blocklist requirements.
\newblock In {\em Proceedings of the 2020 ACM SIGSAC Conference on Computer and Communications Security}, 2020.

\bibitem{SMS2017_Trepte}
Sabine Trepte, Leonard Reinecke, Nicole~B. Ellison, Oliver Quiring, Mike~Z. Yao, and Marc Ziegele.
\newblock A cross-cultural perspective on the privacy calculus.
\newblock {\em Social Media + Society}, 3(1), 2017.

\bibitem{WWW2013_Ur}
Blase Ur and Yang Wang.
\newblock A cross-cultural framework for protecting user privacy in online social media.
\newblock In {\em Proceedings of the 22nd International Conference on World Wide Web}, 2013.

\bibitem{CCS2019_Utz}
Christine Utz, Martin Degeling, Sascha Fahl, Florian Schaub, and Thorsten Holz.
\newblock ({U}n)informed consent: Studying {GDPR} consent notices in the field.
\newblock In {\em Proceedings of the 2019 ACM SIGSAC conference on computer and communications security}, 2019.

\bibitem{SEC2019_Van}
Amber Van Der~Heijden and Luca Allodi.
\newblock Cognitive triaging of phishing attacks.
\newblock In {\em Proceedings of the 28th USENIX Security Symposium}, 2019.

\bibitem{replication_example3}
Melanie Volkamer, Andreas Gutmann, Karen Renaud, Paul Gerber, and Peter Mayer.
\newblock Replication study: A cross-country field observation study of real world {PIN} usage at {ATM}s and in various electronic payment scenarios.
\newblock In {\em Proceedings of the 14th Symposium on Usable Privacy and Security}, 2018.

\bibitem{SP2022_Warford}
Noel Warford, Tara Matthews, Kaitlyn Yang, Omer Akgul, Sunny Consolvo, Patrick~Gage Kelley, Nathan Malkin, Michelle~L Mazurek, Manya Sleeper, and Kurt Thomas.
\newblock {S}o{K}: A framework for unifying at-risk user research.
\newblock In {\em Proceedings of the 43rd IEEE Symposium on Security and Privacy}, 2022.

\bibitem{SOUPS2021_Wash}
Rick Wash, Norbert Nthala, and Emilee Rader.
\newblock Knowledge and capabilities that non-expert users bring to phishing detection.
\newblock In {\em Proceedings of the 17th Symposium on Usable Privacy and Security}, 2021.

\bibitem{SOUPS2015_Wash}
Rick Wash and Emilee Rader.
\newblock Too much knowledge? {S}ecurity beliefs and protective behaviors among {U}nited {S}tates internet users.
\newblock In {\em Proceedings of the 11th Symposium On Usable Privacy and Security}, 2015.

\bibitem{TOCHI2015_Watson}
Jason Watson, Heather~Richter Lipford, and Andrew Besmer.
\newblock Mapping user preference to privacy default settings.
\newblock {\em ACM Transactions on Computer-Human Interaction}, 22(6):1--20, 2015.

\bibitem{271563}
Alma Whitten and J.~D. Tygar.
\newblock Why {J}ohnny can't encrypt: A usability evaluation of {PGP} 5.0.
\newblock In {\em Proceedings of the 8th USENIX Security Symposium}, 1999.

\bibitem{SP2022_Wu}
Yuxi Wu, W.~Keith Edwards, and Sauvik Das.
\newblock {S}o{K}: Social cybersecurity.
\newblock In {\em Proceedings of the 43rd IEEE Symposium on Security and Privacy}, 2022.

\bibitem{democratic_privacy}
Volker Wulf, Dave Randall, Konstantin Aal, and Markus Rohde.
\newblock The personal is the political: Internet filtering and counter appropriation in the {I}slamic {R}epublic of {I}ran.
\newblock {\em Computer Supported Cooperative Work}, 31(2):373–409, 2022.

\bibitem{SP2004_Yan}
Jeff Yan, Alan Blackwell, Ross Anderson, and Alasdair Grant.
\newblock Password memorability and security: Empirical results.
\newblock {\em IEEE Security \& privacy}, 2(5):25--31, 2004.

\bibitem{CHI2020_Zou}
Yixin Zou, Kevin Roundy, Acar Tamersoy, Saurabh Shintre, Johann Roturier, and Florian Schaub.
\newblock Examining the adoption and abandonment of security, privacy, and identity theft protection practices.
\newblock In {\em Proceedings of the 2020 CHI Conference on Human Factors in Computing Systems}, 2020.

\end{thebibliography}

\bigskip
\appendix
\noindent
{\Large{\bf Appendix}}

\section{Our Detailed Screening and Coding Rules}
\label{section:appendix_codingrule}

\subsection{Screening with Search Queries}
\label{section:appendix_screening_detailed}

In Phase--2 of our review process, we filtered papers with the search queries as described in Section~\ref{section:Review Process}. 
We initially considered utilizing the search function of digital libraries but decided not to because we found that the search function of the ACM Digital Library failed to function accurately. 
Specifically, it missed some papers that meet the criteria. 
For the conferences that published PDF files on their websites, we downloaded the files with at least a 10-second interval. 
For the conferences whose papers are stored in digital libraries, we manually downloaded PDF files due to bot prohibition. Then we converted the PDF data into the text data using PDFminer.six~\cite{pdfminer} and ran the query for each paper. 
We preliminarily confirmed that our search query and script matched our criteria. 
Specifically, we ran our script for all accepted papers of a certain conference that we had fully reviewed and found that our script did not exclude papers that meet our criteria.

\subsection{Lists of Reviewed Papers}
\label{section:appendix_papers_list}

Table~\ref{table:papers_list} lists UPS papers we reviewed in this study.
As a result of our paper selection process described in Section~\ref{section:Review Process}, 715 papers were subject to detailed review.

\subsection{Coding of Participants' Countries}
\label{section:appendix_codingrule_country}

\begin{figure*}[h]
 \begin{center}
 \includegraphics[width=0.9\linewidth]{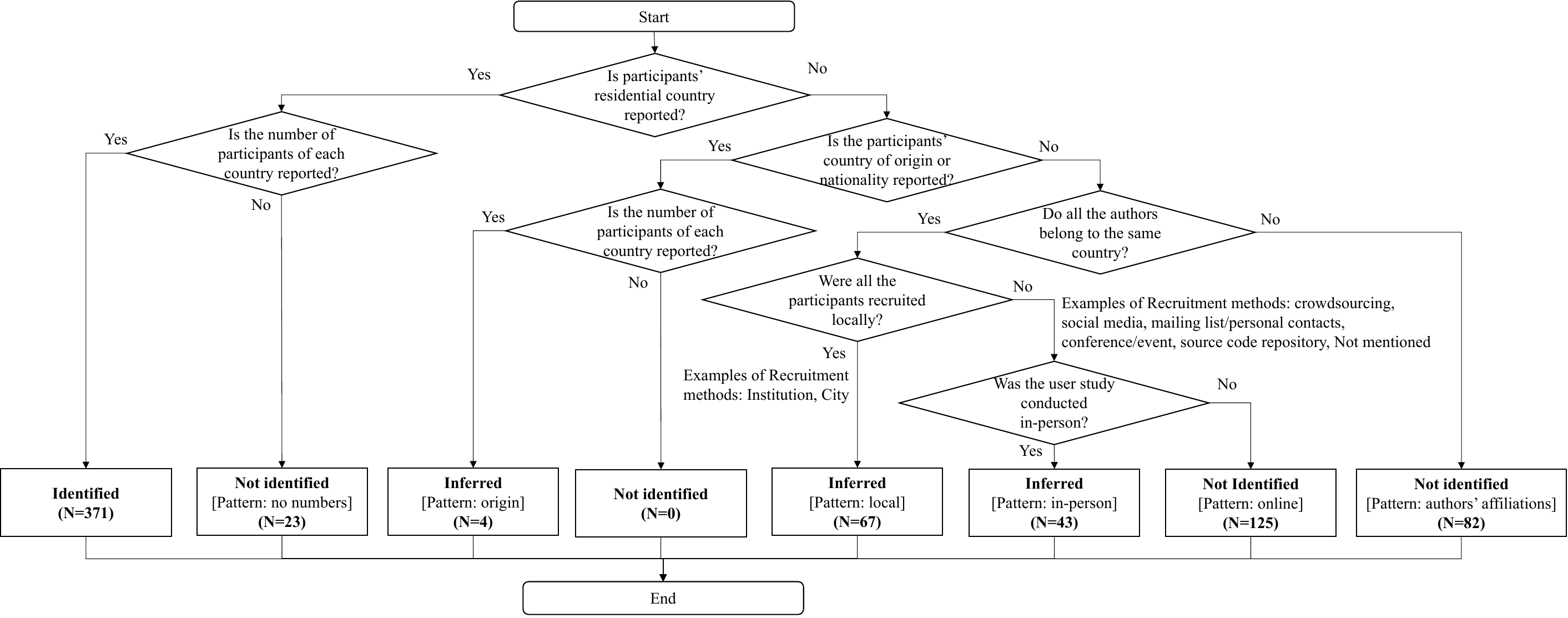}
 \end{center}
 \caption{Flowchart to identify or infer participants' countries.}
 \label{Figs:flow-chart}
\end{figure*}

Figure~\ref{Figs:flow-chart} presents a flowchart of for identifying the participants' countries.
The participant sample in each user study was divided into ``Identified,'' ``Inferred,'' or ``Not identified.''
We found many papers that provided unclear and insufficient descriptions.

\subsection{Coding of Research Topics}
\label{section:appendix_codingrule_topic}

\begin{table}[t] 
\caption{Simplified codebook for analysis of research topics.}
\label{table:Codebook-topics}
\hbox to\hsize{\hfil
\footnotesize{
\begin{tabular}{l|l}\hline
Research topic & Representative description and example work\\
\hline \hline
Overall & Security and privacy practices~\cite{replication_example4} \\
\hline
Access control and & Acceptability for data collection scenarios~\cite{10.1145/3411764.3445122} \\
privacy preference & Permission design~\cite{263788} \\
\hline
Authentication & Password generation and password policies~\cite{10.1145/3372297.3417882}  \\
               & Support tool (e.g., password manager)~\cite{238303} \\
\hline
Vuln./malware/ & Secure software development practices~\cite{10.1145/3313831.3376791} \\
incident response   & Security tool (e.g., static analyzer)~\cite{255682} \\
\hline
Privacy abuse & Intimate partner abuse (IPA)~\cite{10.1145/3025453.3025875} \\
              & Unauthorized access to their accounts~\cite{10.1145/3290605.3300819} \\
\hline
Social engineering & Susceptibility to phishing~\cite{10.1145/3025453.3025831}  \\
      & Phishing-identification support tools~\cite{10.1145/3411764.3445574} \\
\hline
Privacy-enhancing & (Mis)conception of privacy tools~\cite{PETS2021_Story} \\
technologies     & Privacy-enhancing obfuscations~\cite{10.1145/3134702} \\
\hline
\end{tabular}}\hfil}
\end{table}

To classify research topics of UPS papers, we referred to the classification of the research topics in Distler et al.'s work~\cite{TOCHI2021_Distler}.
In that work, the classification of technology (e.g., ``Authentication'') and the classification focusing on user perception and behavior (e.g., ``Security perceptions, attitudes, and behaviors'') were listed together. 
The latter classification can be applied to any UPS papers, e.g., a paper analyzing users' security perception of authentication technology could be classified in either category. 
To disambiguate the classification, we classified research topics only by technology.

Table~\ref{table:Codebook-topics} presents a simplified codebook of research topics and corresponding representative papers.
\begin{itemize} 
\setlength{\itemsep}{2pt}
\setlength{\parskip}{0pt}
\item ``Overall'' includes studies that investigate users' perceptions, attitudes, and behaviors toward a wide range of technologies related to security and privacy, without focusing on specific technologies.
\item ``Access control and privacy preferences'' includes studies that investigate users' privacy concerns, preferences, and practices related to their data and evaluate the usability of privacy control tools.
\item ``Authentication'' includes studies that investigate the security and usability of authentication technologies and the support tools.
\item ``Vulnerability/malware/incident response'' includes studies that investigate response skills and practices regarding vulnerabilities and incidents, and evaluate the usability of the support tools. 
\item ``Privacy abuse'' includes studies that investigate the threats to user privacy, e.g., survivors' defense practices and concerns, and abusers' approaches.
\item ``Social engineering'' includes studies that investigate users' susceptibility to social engineering and defense practices. 
It also includes studies that investigate the effectiveness of the countermeasures (e.g., warnings).
\item ``Privacy-enhancing technologies'' includes studies that investigate users' (mis)conceptions and usage of privacy-enhancing technologies (e.g., VPNs and E2EE). 
\item ``Other'' includes any studies that focus on technologies or threats other than those listed above, e.g., adversarial attacks (e.g., insertion of malicious commands into personal assistants).
\end{itemize}

\section{WEIRD with AND Operators}
\label{section:appendix_WEIRD_with_AND_Operators}

In Sections~\ref{section:results_western}, ~\ref{section:results_educated}, and ~\ref{section:results_industrialized_and_rich}, we have shown that most UPS participant samples are from WEIRD countries using OR as the logical operator.
We are also interested in the percentage of participant samples that are from fully WEIRD countries, i.e., using AND as the logical operator.
Similar to the Linxen study, we set a median for each of the E, I, R, and D distributions of the countries that the UPS papers studied once or more as a cutoff value.
Note that this analysis shows the relatively fully-WEIRD countries among the countries studied in the UPS papers but not the fully-WEIRD countries among all the countries in the world.
Of 791 participant samples, we found that 75.22\% were recruited in fully WEIRD countries, 3.67\% were recruited in Western but not ``EIRD'' countries, 2.53\%  were recruited in non-Western but ``EIRD'' countries (e.g., Israel and Japan), and the remaining 18.58\% were recruited in fully non-WEIRD countries (e.g., India and Bangladesh).

\section{Hofstede's Cultural Dimensions}
\label{section:appendix_Hofstede}

We performed the additional analysis using Hofstede's cultural dimensions framework~\cite{hofstede, hofstede_website}. 
This framework is used for understanding cultural differences and consists of six cultural dimensions, as shown in Table~\ref{table:results_correlation_hofstede}.
Although the framework cannot explain all cultural differences, we believe that these dimensions are relevant to UPS research. 
For example, countries with higher power distance exhibited significantly higher levels of government involvement in regulating information privacy~\cite{CACM1995_Milberg, EM2016_Cockcroft}. 
As for information disclosure on social media, Trepte et al.~\cite{SMS2017_Trepte} focused on the conflicting findings of the prior studies: (1) people from individualistic cultures are more likely to avoid privacy risks, and (2) people from collectivist cultures are more likely to avoid them. 
Their result supported the latter findings.
As for phishing, Butavicius et al. demonstrated that users from individualistic cultures were significantly less likely to be susceptible to phishing emails~\cite{HAISA2017_Butavicius}.

\begin{table}[tbh] 
\caption{Correlations of $\psi_{\scalebox{0.9}[1.0]{\textit{s}}}$ with cultural dimensions.}
\label{table:results_correlation_hofstede}
\hbox to\hsize{\hfil
\footnotesize{
\begin{tabular}{l|r}\hline
\multicolumn{1}{l|}{Indicators} & \multicolumn{1}{c}{$r$} \\\hline\hline
Power distance & -.44***\\\hline
Individualism (vs. collectivism) & .43***\\\hline
Masculinity (vs. femininity) & -.09\\\hline
Uncertainty avoidance & -.08\\\hline
Long-term orientation (vs. short-term) & .13\\\hline
Indulgence (vs. restraint) & .09\\\hline
\end{tabular}}\hfil}
\end{table}

We collected six dimension scores per country~\cite{hofstede_website} and calculated the correlations between countries' {\small$\psi_{\scalebox{0.9}[1.0]{\textit{samples}}}$} and each indicator of a cultural dimension. 
We found that {\small$\psi_{\scalebox{0.9}[1.0]{\textit{samples}}}$} has a negative correlation ($r=$-.44, $p<$.001) with power distance and a positive correlation ($r=$.43, $p<$.001) with individualism.
That is, most participant samples in UPS papers come from individualistic cultures that have equal distributions of power.
Considering the findings of prior studies and our analysis, we highlighted that most UPS papers may have investigated people primarily from countries whose residents are more likely to disclose their information on social media and are less likely to be phished.

\clearpage
\onecolumn


\begin{table}[t] 
\caption{Lists of Reviewed Papers ($N=$715).}
\label{table:papers_list}
\hbox to\hsize{\hfil
\scriptsize{
\begin{tabular}{llp{13cm}}\hline
Conference & Year & Title \\\hline
S\&P & 2017 & SmarPer: Context-Aware and Automatic Runtime-Permissions for Mobile Devices \\
S\&P & 2017 & Cloak and Dagger: From Two Permissions to Complete Control of the UI Feedback Loop \\
S\&P & 2017 & XHOUND: Quantifying the Fingerprintability of Browser Extensions \\
S\&P & 2017 & Comparing the Usability of Cryptographic APIs \\
S\&P & 2017 & An Experimental Security Analysis of an Industrial Robot Controller \\
S\&P & 2017 & The Password Reset MitM Attack \\
S\&P & 2017 & SysPal: System-guided Pattern Locks for Android \\
\hline
\end{tabular}}\hfil}
\end{table}

\begin{table}[t] 
\hbox to\hsize{\hfil
\scriptsize{
\begin{tabular}{llp{13cm}}\hline
Conference & Year & Title \\\hline
SEC & 2020 & A different cup of TI? The added value of commercial threat intelligence \\
SEC & 2020 & Chaperone: Real-time Locking and Loss Prevention for Smartphones \\
SEC & 2020 & Devil’s Whisper: A General Approach for Physical Adversarial Attacks against Commercial Black-box Speech Recognition Devices \\
SEC & 2020 & The Impact of Ad-Blockers on Product Search and Purchase Behavior: A Lab Experiment \\
SEC & 2020 & The Industrial Age of Hacking \\
SEC & 2020 & Adversarial Preprocessing: Understanding and Preventing Image-Scaling Attacks in Machine Learning \\
SEC & 2020 & A Comprehensive Quality Evaluation of Security and Privacy Advice on the Web \\
\hline
\end{tabular}}\hfil}
\end{table}

\begin{table}[t] 
\hbox to\hsize{\hfil
\scriptsize{
\begin{tabular}{llp{13cm}}\hline
Conference & Year & Title \\\hline
CCS & 2019 & (Un)informed Consent: Studying GDPR Consent Notices in the \\
CCS & 2019 & 28 Blinks Later: Tackling Practical Challenges of Eye Movement Biometrics \\
CCS & 2019 & Matched and Mismatched SOCs: A Qualitative Study on Security Operations Center Issues \\
CCS & 2019 & Velody: Nonlinear Vibration Challenge-Response for Resilient User Authentication \\
CCS & 2019 & The Catcher in the Field: A Fieldprint based Spoofing Detection for Text-Independent Speaker Verification \\
CCS & 2019 & You Are Who You Appear to Be A Longitudinal Study of Domain Impersonation in TLS Certificates \\
CCS & 2019 & Oh, the Places You’ve Been! User Reactions to Longitudinal Transparency About Third-Party Web Tracking and Inferencing \\
\hline
\end{tabular}}\hfil}
\end{table}

\begin{table}[t] 
\hbox to\hsize{\hfil
\scriptsize{
\begin{tabular}{llp{13cm}}\hline
Conference & Year & Title \\\hline
PETS & 2020 & Explaining the Technology Use Behavior of Privacy-Enhancing Technologies: The Case of Tor and JonDonym \\
PETS & 2020 & SPy: Car Steering Reveals Your Trip Route! \\
PETS & 2020 & Listen Only When Spoken To: Interpersonal Communication Cues as Smart Speaker Privacy Controls \\
PETS & 2020 & Privacy at a Glance: The User-Centric Design of Glanceable Data Exposure Visualizations \\
PETS & 2020 & Smart Devices in Airbnbs: Considering Privacy and Security for both Guests and Hosts \\
PETS & 2020 & Multiple Purposes, Multiple Problems: A User Study of Consent Dialogs after GDPR \\
PETS & 2020 & Long-Term Observation on Browser Fingerprinting: Users’ Trackability and Perspective \\
\hline
\end{tabular}}\hfil}
\end{table}

\begin{table}[t] 
\hbox to\hsize{\hfil
\scriptsize{
\begin{tabular}{llp{13cm}}\hline
Conference & Year & Title \\\hline
CHI & 2018 & Unpacking Perceptions of Data-Driven Inferences Underlying Online Targeting and Personalization \\
CHI & 2018 & An Experience Sampling Study of User Reactions to Browser Warnings in the Field \\
CHI & 2018 & “This App Would Like to Use Your Current Location to Better Serve You”: Importance of User Assent and System Transparency in Personalized Mobile Services \\
CHI & 2018 & Forgotten But Not Gone: Identifying the Need for Longitudinal Data Management in Cloud Storage \\
CHI & 2018 & Leveraging Semantic Transformation to Investigate Password Habits and Their Causes \\
CHI & 2018 & “A Stalker’s Paradise”: How Intimate Partner Abusers Exploit Technology \\
CHI & 2019 & “They Don’t Leave Us Alone Anywhere We Go”: Gender and Digital Abuse in South Asia \\
\hline
\end{tabular}}\hfil}
\end{table}

\begin{table}[t] 
\hbox to\hsize{\hfil
\scriptsize{
\begin{tabular}{llp{13cm}}\hline
Conference & Year & Title \\\hline
CHI & 2020 & Improving the Usability and UX of the Swiss Internet Voting Interface \\
CHI & 2020 & On Conducting Security Developer Studies with CS Students: Examining a Password-Storage Study with CS Students, Freelancers, and Company Developers \\
CHI & 2020 & “I’m hoping they’re an ethical company that won’t do anything that I’ll regret”: Users’ Perceptions of At-home DNA Testing Companies \\
CHI & 2020 & Online Privacy Heuristics that Predict Information Disclosure  \\
CHI & 2020 & Understanding Cybersecurity Practices in Emergency Departments \\
CHI & 2021 & “You Gota Watch What You Say”: Surveillance of Communication with Incarcerated People \\
CHI & 2021 & Understanding the Security and Privacy Advice Given to Black Lives Matter Protesters \\
\hline
\end{tabular}}\hfil}
\end{table}

\begin{table}[t] 
\hbox to\hsize{\hfil
\scriptsize{
\begin{tabular}{llp{13cm}}\hline
Conference & Year & Title \\\hline
CSCW & 2018 & “Our Privacy Needs to be Protected at All Costs”: Crowd Workers’ Privacy Experiences on Amazon Mechanical Turk \\
CSCW & 2018 & The Influence of Friends and Experts on Privacy Decision Making in IoT Scenarios \\
CSCW & 2018 & Going Beyond Obscurity: Organizational Approaches to Data Anonymization \\
CSCW & 2018 & Alexa, Are You Listening? Privacy Perceptions, Concerns and Privacy-seeking Behaviors with Smart Speakers \\
CSCW & 2018 & Tell Me Before You Stream Me: Managing Information Disclosure in Video Game Live Streaming \\
CSCW & 2018 & When SNS Privacy Settings Become Granular: Investigating Users’ Choices, Rationales, and Influences on Their Social Experience \\
CSCW & 2018 & Distinguishing Group Privacy From Personal Privacy: The Effect of Group Inference Technologies on Privacy Perceptions and Behaviors \\
\hline
\end{tabular}}\hfil}
\end{table}

\begin{table}[t] 
\hbox to\hsize{\hfil
\scriptsize{
\begin{tabular}{llp{13cm}}\hline
Conference & Year & Title \\\hline
SOUPS & 2018 & “We make it a big deal in the company”: Security Mindsets in Organizations that Develop Cryptographic Products \\
SOUPS & 2018 & “It’s Scary...It’s Confusing...It’s Dull”: How Cybersecurity Advocates Overcome Negative Perceptions of Security \\
SOUPS & 2018 & Programming Experience Might Not Help in Comprehending Obfuscated Source Code Efficiently \\
SOUPS & 2018 & Data Breaches: User Comprehension, Expectations, and Concerns with Handling Exposed Data \\
SOUPS & 2018 & Characterizing the Use of Browser-Based Blocking Extensions To Prevent Online Tracking \\
SOUPS & 2018 & Online Privacy and Aging of Digital Artifacts \\
SOUPS & 2018 & “If I press delete, it’s gone” - User Understanding of Online Data Deletion and Expiration \\
\hline
\end{tabular}}\hfil}
\end{table}

\begin{table}[t] 
\hbox to\hsize{\hfil
\scriptsize{
\begin{tabular}{llp{13cm}}\hline
Conference & Year & Title \\\hline
SOUPS & 2021 & Concerned but Ineffective: User Perceptions, Methods, and Challenges when Sanitizing Old Devices for Disposal \\
SOUPS & 2021 & PowerCut and Obfuscator: An Exploration of the Design Space for Privacy-Preserving Interventions for Smart Speakers \\
SOUPS & 2021 & Towards Usable and Secure Location-based Smartphone Authentication \\
SOUPS & 2021 & Virtual Classrooms and Real Harms: Remote Learning at U.S. Universities \\
SOUPS & 2021 & Code Reviewing as Methodology for Online Security Studies with Developers – A Case Study with Freelancers on Password Storage \\
SOUPS & 2021 & Understanding Privacy Attitudes and Concerns Towards Remote Communications During the COVID-19 Pandemic \\
SOUPS & 2021 & Benefits and Drawbacks of Adopting a Secure Programming Language: Rust as a Case Study \\
\hline
\end{tabular}}\hfil}
\end{table}

\begin{table}[t] 
\hbox to\hsize{\hfil
\scriptsize{
\begin{tabular}{llp{13cm}}\hline
Conference & Year & Title \\\hline
EuroUSEC & 2017 & Personalized Security Messaging: Nudges for Compliance with Browser Warnings \\
EuroUSEC & 2017 & Effects of information security risk visualization on managerial decision making \\
EuroUSEC & 2017 & Information Leakage through Mobile Motion Sensors: User Awareness and Concerns \\
EuroUSEC & 2017 & I Do and I Understand. Not Yet True for Security APIs. So Sad \\
EuroUSEC & 2017 & Can Johnny Build a Protocol? Co-ordinating developer and user intentions for privacy-enhanced secure messaging protocols \\
EuroUSEC & 2017 & Pico in the Wild: Replacing Passwords, One Site at a Time \\
EuroUSEC & 2017 & What is a Secure Email? \\
\hline
\end{tabular}}\hfil}
\end{table}

\end{document}